\documentclass[a4paper,11pt]{article}
\usepackage[utf8]{inputenc}

\usepackage{amsmath, amsthm}
\usepackage{amsfonts}
\usepackage{multirow}
\usepackage[english]{babel}
\usepackage[utf8]{inputenc}
\usepackage{geometry}
\usepackage{graphicx} % used to insert the figure
\usepackage{xspace}
\title{Boundary Element and Finite Element Coupling \\ for Aeroacoustics Simulations}
\author{Nolwenn Balin$^1$, Fabien Casenave$^{1,2}$, Fran\c cois Dubois$^3$, \\
Eric Duceau$^1$, Stefan Duprey$^{1,4}$, Isabelle Terrasse$^{1}$\\~\\
\small $^1$ Airbus Group Innovations, 12 Rue Pasteur, 92150 Suresnes, France\\
\small $^2$ Universit\'{e} Paris-Est, CERMICS (ENPC), 6-8 Avenue Blaise Pascal, Cit\'{e} Descartes,\\
\small F-77455 Marne-la-Vall\'{e}e, France\\
\small $^3$ CNAM, Laboratoire LMSSC, 292 rue Saint-Martin, 75141 Paris Cedex 03, France\\
\small $^4$ Institut Elie Cartan, Universit\'e Henri Poincar\'e, 24-30 rue Lionnois,  
54003 Nancy Cedex, France 
}

\def\N {\mathbb{N}}
\def\Z {\mathbb{Z}}

\newcommand{\nc}{\newcommand}

\nc{\dsp}{\displaystyle}
\nc{\y}{\mathbf{y}}
\nc{\dt}{\mathit{\Delta t}}
\nc{\dx}{\mathit{\Delta x}}
\nc{\dy}{\mathit{\Delta y}}
\nc{\PRe}{\Re e}

\nc{\ovra}{\overrightarrow}
\nc{\ud}{\mathrm{d}}

\nc{\xt}{(\vec{x},t)}
\nc{\rz}{(r,z)}
\nc{\rzte}{(r,z,\theta)}
\nc{\nablav}{\ovra{\nabla}}
\nc{\x}{\vec{x}}
\nc{\n}{\vec{n}}
\nc{\V}{\vec{v}}
\nc{\vinf}{v_{\infty}}
\nc{\ainf}{a_{\infty}}
\nc{\pas}{{\psi_a^*}}
\nc{\pam}{{\psi_{am}^*}}
\nc{\Vas}{\vec{v_a^*}}
\nc{\kos}{{k_0}}

\nc{\er}{\vec{e}_r}
\nc{\ez}{\vec{e}_z}
\nc{\ete}{\vec{e}_{\theta}}

\nc{\mn}{\iota}
\nc{\mtimesn}{\iota}
\nc{\pq}{\kappa}
\nc{\ptimesq}{\kappa}

\nc{\trinf}[1]{{\widetilde{#1}}{}}
\nc{\trM}[1]  {{\widetilde{#1}^{M}}{}}
\nc{\DtN}{{\Lambda}}

\renewcommand{\vec}[1]{\boldsymbol{#1}}
\renewcommand{\bar}[1]{\overline{#1}}
\nc{\unvec}[1]{\hat{#1}}

\nc{\grad}{\vec{\nabla}}

\nc{\inc}{{\textrm{inc}}}
\nc{\diffr}{{\textrm{diff}}}

\nc{\PGtransf}{{Prandtl--Glauert transformation}\xspace}
\nc{\actipole}{{ACTIPOLE}\xspace}

\nc{\todo}[1]{\textit{\textbf{todo: #1 !}}}

\newtheorem{definition}{Definition}[section]

\newtheorem{property}{Property}[section]

\newtheorem{remark}{Remark}[section]

\nc{\indices}[4]{
\scriptsize
\begin{array}{c}
1\leq #1\leq #2\\
1\leq #3\leq #4
\end{array}
}

\nc{\indiceslight}[2]{
\scriptsize
\begin{array}{c}
#1\\
#2
\end{array}
}

\begin{document}

\maketitle

\begin{abstract}
We consider the scattering of acoustic perturbations in a presence of a flow. We suppose that the space can be split into
a zone where the flow is uniform and a zone where the flow is potential.
In the first zone, we apply a Prandtl--Glauert
transformation to recover the Helmholtz equation. The well-known setting of boundary element method for the
Helmholtz equation is available. In the second zone, the flow quantities are space dependent, we have to
consider a local resolution, namely the finite element method. Herein, we carry out the coupling of these two methods
and present various applications and validation test cases.
The source term is given through the decomposition
of an incident acoustic field on a section of the computational domain's boundary.
\end{abstract}

\section{Introduction}

\indent Acoustics is a well known science and the basics
mechanical and thermodynamical notions are well
understood since the 19th century as shown  {\it e.g.}
with the classical books of Lord Rayleigh \cite{Ra77}.  
For a modern presentation of various aspects of this science, 
we refer to Morse and Ingard \cite{MI68}  
and to the contribution of  Bruneau {\it et al.}   \cite{Br06}.
Acoustics can be presented with   temporal or 
harmonic dynamics. In the first case, acoustics can be viewed as 
an hyperbolic problem and in the second, the Helmholtz equation
plays a central role. 
 
With direct numerical  time integration, 
finite differences are naturally popular.
Even if it has not be created for acoustics applications, the ``Marker And Cell''
method with  staggered grids of Harlow and Welch \cite{HWSD66}  can be used 
very easily in acoustics. 
We refer also to the 
pioneering work of Virieux for geophysics applications \cite{Vi86}.
A finite difference method
uses a finite grid in a domain of finite size.
How to express that waves can go outside the computational domain 
without reflection~? 
One possible solution is to derive appropriate absorbing  boundary
conditions (see {\it e.g.} the book of   Taflove summarized  \cite {Ta95}).   
Another possibility is to add a layer of absorbing material.
Efficient absorbing layers have been first proposed 
 for the vectorial wave equation (Maxwell in electromagnetism)
by  B\'erenger \cite{Be94}, then applied in acoustics 
(scalar equation) by  Abarbanel {\it et al.}  \cite{AGH99}.
It was adapted by our group for advective acoustics 
and staggered grids \cite{DDMT}. However, the adaptation of
cartesian finite differences to   complex industrial
 geometries is a very difficult task and other 
numerical methods have been developed in order 
to guarantee this flexibility. 
The finite element method is the most 
popular in this direction. 
We refer to the  fundamental book of Zienkiewicz \cite{Zi67}
essentially for structural mechanics applications
and to  Craggs \cite{Cr73}  for the acoustics applications. 
A rigorous mathematical analysis  of the method with
Hilbertian mathematical methods is proposed in the book 
of  Ciarlet \cite{Ci78}. The main advantages of these so-called volume methods
is the possibility to deal with space dependent media of propagation.
 
When the medium of propagation is uniform, the opportunity to 
represent the field as an  integral representation  over a surface of 
some data on the boundary of the radiating object makes natural
so-called integral methods. The unknown is a field simply 
located on a finite surface and the three-dimensional field
couples all the degrees of freedom on the surface. 
The difficulty is to take into account the fact that waves are 
radiating from finite distance  towards infinity.
The radiation Sommerfeld condition solves this problem
and  expresses that no wave is coming from infinity \cite{So12}. 
The adaptation of these ideas to integral methods for 
 exterior boundary-value problems for  Helmholtz equation 
have been discussed among others by  Schenck \cite{Sc68} 
and Burton and Miller \cite {BM71}.
For a rigorous mathematical analysis, we refer to 
 N\'ed\'elec \cite{Ne01}. The main advantages of these so-called integral methods
is the possibility to deal with large geometries. In particular,
in the case of the scattering by two objects,
the size of the numerical problem does not depend on the distance between the
objects. Besides, we have access to the scattered field at any point of the space.

A natural idea to profit from the advantages of volume and integral methods is the coupling 
between boundary and finite element methods. The fundamental mathematical  work is due
to Zienkiewicz, Kelly, and Bettess \cite{Zi67}, Johnson and  N\'ed\'elec \cite{JN80}, and
Costabel \cite{Costabel}. The thesis of Levillain \cite {Le91} gives
the first numerical applications for  Maxwell equations. We refer to Bielak and Mac Camy \cite{BM91} 
for fluid - solid coupling. For other modern developments, we refer to the work of 
Abboud {\it et al.} (see {\it e.g.} \cite{AJRT11}).

Integral methods have been implemented using Boundary Element Method (BEM) at former Airbus Research Center at Suresnes and Toulouse (France) since 2001  \cite{justif_actipole_2011,delnevo} and used by Airbus  first for computation of radiated noise outside air inlet than exhaust with the assumption of uniform flow on whole domain \cite{delnevo2005aiaa}.

Several improvements have then been conducted by our team in two main ways : (i) increasing the number of unknowns - leading to the computation of the acoustic propagation at higher frequencies - using efficient numerical method such as the Fast Multipole Method~\cite{sylvand2002phd, 2006:ParallelSolverLargeProblems}; (ii) increasing the complexity of the flow to deal with more realistic problems. 
First results for coupled BEM-FEM with a uniform flow at infinity and potential flow assumption close to the scattering object have been obtained in an axisymmetric configuration in~\cite{duprey}.
Mathematical framework in the 3D case have been presented in~\cite{casenave}. In this contribution, we present an extension of the previous works in a more general context closer to applications  and taking into account realistic air inlets: our main objective is to perform a global simulation from a given Mach number in the duct to a different Mach number in far field with a transitional potential flow between the two zones.

The model coupled problem is presented in Section \ref{sec:modelpb}. This problem is transformed by the \PGtransf,
in order to recover the classical Helmholtz equation in the area where the flow is uniform in Section \ref{sec:lorentz}, leading
to a transformed coupled weak formulation. The finite dimensional linear system is presented in Section \ref{sec:meth_num}, and
numerical studies are carried out in Section \ref{sec:numericalresults}.

\section{Definition of the model problem}

\label{sec:modelpb}

\subsection{Context and geometry}

\begin{figure}[htbp]
 \centering
 \includegraphics[width=0.6\textwidth]{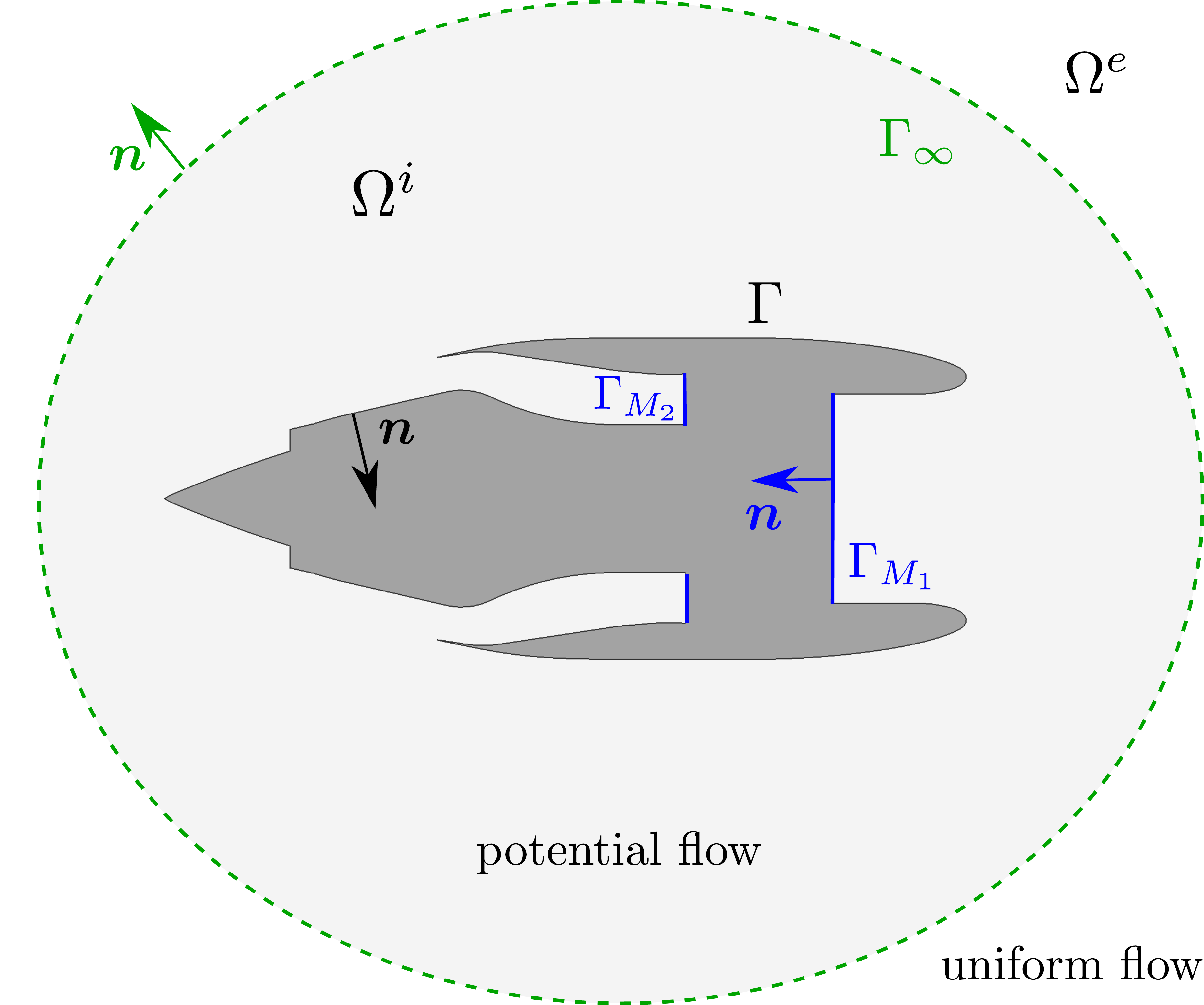}
 % dessin_pbmodele_engine.pdf: 1059x884 pixel, 72dpi, 37.36x31.19 cm, bb=0 0 1059 884
 \caption{Model problem (without representation of the fan spinner)}
 \label{fig:model_pb_engine}
\end{figure}

The objective of the current work is the computation of the acoustic field generated by
a turboreactor engine in flight condition, especially in take-off and landing phases. We will consider the model problem presented in Figure \ref{fig:model_pb_engine}, where the other parts of the aircraft (engine pylon, wings, fuselage) are not modeled.

The considered acoustic sources are the inward and outward fans noise.
The fan noise frequency spectrum is characterized by some harmonic peaks at frequencies that are multiple of the rotational frequency of the blades.
For simplicity, we consider a single pulsation $\omega_0$.
Moreover, the fans are located inside a duct that is relatively deep compared to its width. For these reasons, it is classical
to model the duct by semi-infinite cylinders, and represent the acoustic sources on modal bases functions defined at the bases
of these cylinders, called modal surfaces.
As illustrated on Figure \ref{fig:model_pb_engine}, two modal surfaces $\Gamma_{M_1}$ and $\Gamma_{M_2}$
are defined. The other part of the engine boundary is rigid and noted $\Gamma$.

The complexity of the flow depends on the distance from the engine. For low Mach values, the flow can be decomposed into three areas with different flow conditions: an uniform flow far from the engine, a turbulent flow (often non-linear) due to the fans and the boundary layers, and a potential flow in-between. 
We use this decomposition in our model, except for the turbulent area.%: the impact of the boundary layers are neglected. 

The impact of the viscosity effect and in particular the boundary layers is neglected in this
contribution. This assumption is classical in the community of industrial acoustics. 
In air intake cases (see {\it e.g.} Lidoine \cite{Li02}), there is no flow separation. 
The boundary layer remains very thin compared to the wavelength of interest and is not taken into account for the acoustic point of view.
The problem of thin boundary layers in the duct flow has been studied 
by  Eversman \cite {Ev73}. This question has been emphasized after the work of Myers \cite{My80}. The Myers boundary condition has been improved by
Brambley \cite{Br11}.
For an ejection duct, the situation is more complicated because anyway the flow is far from potential.
In particular, a mixing layer containing different jets is present. Moreover, its thickness is small near the ejection 
compared to the characteristics lengths of the problem. As a consequence, it is reasonable to assume that the sound has a negligible influence on the flow.

The interior part of the engine is also supposed to be behind the modal surfaces. We then consider only two domains: $\Omega^i$ with a potential flow and $\Omega^e$ with an uniform flow. 
The flow is an input to the problem.
Matching conditions on the flow are supposed at the interface $\Gamma_\infty$ between these two domains. Moreover, at $\Gamma_{M_1}$ and $\Gamma_{M_2}$, the flow is supposed uniform and orthogonal to the surface. It it also tangent to $\Gamma$.

\subsection{Coupled problem}

\subsubsection{Convected Helmholtz equation}

We note $c_0$ the celerity, $\rho_0$ the density, $\vec{M_0}$ the Mach vector of the fluid.
These quantities depend on the position in $\Omega^i$:%$\forall \vec{x}\in\Omega^e \cup \Gamma_\infty$, 
\begin{align*}
&\forall \vec{x}\in\Omega^e \cup \Gamma_\infty: & c_0(\vec{x}) &= c_\infty, &  \rho_0(\vec{x}) &= \rho_\infty, &  \vec{M_0}(\vec{x}) & = \vec{M_\infty}, \\
&\forall \vec{x}\in\Gamma_M:                    & c_0(\vec{x}) &= c_M,      &  \rho_0(\vec{x}) &= \rho_M, &  \vec{M_0}(\vec{x}) & = \vec{M_M}.
\end{align*}

We note $\vec{v}$ and $p$ the acoustic velocity and pressure.

In the interior domain $\Omega^i$, the convective flow is supposed to be subsonic, stationary, non viscous, isentropic, and irrotational.
Moreover, the acoustic effects are considered to be a first order perturbation of this flow. With these assumptions, the acoustic velocity $\vec{v}$
is potential. Hence, there exists an acoustic potential, $\varphi$, such that $\vec{v}=\vec{\nabla}{\varphi}$.
We note by $\varphi^e$ and $\varphi^i$ the acoustic potential respectively restricted to $\Omega^e$ and $\Omega^i$. 

The physical quantities are associated with complex quantities with the following convention on, for instance, the acoustic potential:
$\varphi\leftrightarrow\Re\left(\varphi \exp({-i\omega_0 t})\right)$. %, where $\omega$ is the pulsation of the source.
The same notation is taken for the physical quantity and its complex counterpart.
In what follows, we always refer to the complex quantities.
The wavenumber depends on the position in $\Omega^i$: 
\begin{align*}
k_0(\vec{x}) &= k_\infty := \frac{\omega_0}{c_\infty}\quad \forall \vec{x}\in\Omega^e \cup \Gamma_\infty, & 
k_0(\vec{x}) &= k_M  := \frac{\omega_0}{c_M}\quad \forall \vec{x}\in\Gamma_M.
\end{align*}

For simplicity, only one modal surface $\Gamma_M$ is considered in the model problem instead of $\Gamma_{M_1}$ and $\Gamma_{M_2}$ as presented on Figure \ref{fig:model_pb_engine}.
Following \cite[p.259 eq.F27]{introacous}, and making use of the irrotationality of the carrier flow, the linearization of the Euler equations leads to
\begin{equation}
\mathcal{H}(\varphi)=0 \quad\text{in } \Omega^i\cup\Gamma_\infty\cup\Omega^e,
\label{eq:globalequation_general_problem}
\end{equation}
where $\mathcal{H}(\varphi):= \rho_0\left(k_0^2\varphi+i k_0 \vec{M_0}\cdot\vec{\nabla}{\varphi}\right)+{\rm div}\left[\rho_0\left(\vec{\nabla}{\varphi}
-\left(\vec{M_0}\cdot\vec{\nabla}{\varphi}\right)\vec{M_0}+i k_0 \varphi \vec{M_0}\right)\right]$.
This is the convected Helmholtz equation.
We assume that $\Gamma$ reflects perfectly the acoustic perturbations, yielding

\begin{equation}
\vec\nabla{\varphi}\cdot\vec{n}=0\quad\text{on }\Gamma.
\end{equation}
We now explain how the problem in~$\Omega^i$ is coupled to a problem in~$\Omega^e$ and a problem beyond $\Gamma_M$ by means of boundary conditions~on $\Gamma_\infty$ and~$\Gamma_{M}$.

\subsubsection{Coupling by means of boundary conditions}

To focus on the coupling between the problem in~$\Omega^i$ and the one in~$\Omega^e$, consider a test case like the one Figure~\ref{fig:model_pb_engine}, where some boundary
condition is enforced on $\Gamma_M$:
\begin{equation}
\left\{
\begin{aligned}
\mathcal{H}(\varphi) &=0, &&\textnormal{~in }\Omega^i\cup\Gamma_\infty\cup\Omega^e,\\
\vec{\nabla}\varphi\cdot\vec{n}&=0,&&\textnormal{~on }\Gamma,\\
\varphi&=g,&&\textnormal{~on }\Gamma_M,\\
\textnormal{Sommerfeld radiation condition},&&
\end{aligned}
\right.
\label{eq:pb_no_modal}
\end{equation}
where a Sommerfeld-like radiation condition is enforced to ensure uniqueness of problem by selecting outgoing scattered waves and where $g$ is some nonzero function defined
on $\Gamma_M$.
Consider the following problem
\begin{equation}
\left\{
\begin{aligned}
\mathcal{H}(\varphi) &=0, &&\textnormal{~in }\Omega^i,\\
\mathcal{H}(\varphi) &=0, &&\textnormal{~in }\Omega^e,\\
\left[\varphi\right]_{\Gamma_\infty}&=0,&&\textnormal{~on }\Gamma_\infty,\\
\left[\left(\vec{\nabla}{\varphi}
-\left(\vec{M_\infty}\cdot\vec{\nabla}{\varphi}\right)\vec{M_\infty}+i k_\infty \varphi \vec{M_\infty}\right)\cdot\vec{n}\right]_{\Gamma_\infty}&=0,&&\textnormal{~on }\Gamma_\infty,\\
\vec{\nabla}\varphi\cdot\vec{n}&=0,&&\textnormal{~on }\Gamma,\\
\varphi&=g,&&\textnormal{~on }\Gamma_M,\\
\textnormal{Sommerfeld-like radiation condition},&&
\end{aligned}
\right.
\label{eq:pb_no_modal2}
\end{equation}
where $[\cdot]_X$ denotes the jump of a quantity across a surface $X$.
\begin{property}
Problems~\ref{eq:pb_no_modal} and~\ref{eq:pb_no_modal2} are equivalent.
\end{property}
\begin{proof}
If $\varphi$ solves~\eqref{eq:pb_no_modal}, it is clear that $\varphi$ solves~\eqref{eq:pb_no_modal2}. Conversely, let $\varphi$ defined in $\Omega^i\cup\Omega^e$ such that $\varphi$
verifies~\eqref{eq:pb_no_modal2}. From~\cite[Lemma~4.19]{mclean}, since $\varphi$ verifies lines 1,~2,~3,~4 of~\eqref{eq:pb_no_modal2}, then $\varphi$ verifies line 1
of~\eqref{eq:pb_no_modal}, which finishes the proof.
\end{proof}

Using the transmission conditions of problem~\eqref{eq:pb_no_modal2} on~$\Gamma_\infty$, we will write problem~\eqref{eq:pb_no_modal} in the form of a problem on $\varphi^i$ and
a boundary condition on~$\Gamma_\infty$. Consider the following problem in $\Omega^e$
\begin{equation}
\left\{
\begin{aligned}
\mathcal{H}(\varphi^e) &=0, &&\textnormal{~in }\Omega^e,\\
\varphi^e&=F,&&\textnormal{~on }\Gamma_\infty,\\
\textnormal{Sommerfeld-like radiation condition},&&
\end{aligned}
\right.
\label{eq:pb_no_modal3}
\end{equation}
where $F$ is some nonzero function on $\Gamma_\infty$. With some regularity hypothesis on $\Omega^e$ and $F$, problem~\eqref{eq:pb_no_modal3} has a unique solution in
$H^1_{\rm loc}(\Omega^e)$, where $H^1_{\rm loc}(\Omega^e):=\{u\in H^1(K),\forall K\subset\Omega^e~\textnormal{compact}\}$,~\cite[Theorem 9.11]{mclean}.
\begin{remark}
\cite[Theorem 9.11]{mclean} deals with the classical Helmholtz equations. We will see in Section~\ref{sec:lorentz} that the classical Helmholtz equation
can be obtained from the convected Helmholtz equation applying a certain transformation. In particular, the solution functions are changed by an injective transformation,
ensuring existence and uniqueness of problem~\eqref{eq:pb_no_modal3}.
\end{remark}
Thus, it is possible to define the operator that maps $F$ onto $\left(\vec{\nabla}{\varphi^e}
-\left(\vec{M_\infty}\cdot\vec{\nabla}{\varphi^e}\right)\vec{M_\infty}+i k_\infty \varphi^e \vec{M_\infty}\right)|_{\Gamma_\infty}\cdot\vec{n}$,
where $\varphi^e$ is the solution of problem~\eqref{eq:pb_no_modal3}. This operator is called the Dirichlet-to-Neumann operator associated with
the exterior problem~\eqref{eq:pb_no_modal3}, and is denoted by $\DtN_{\infty}$. Hence, problem~\eqref{eq:pb_no_modal} is written
\begin{equation}
\left\{
\begin{aligned}
\mathcal{H}(\varphi^i) &=0, &&\textnormal{~in }\Omega^i,\\
\left(\vec{\nabla}{\varphi^i}-\left(\vec{M_\infty}\cdot\vec{\nabla}{\varphi^i}\right)\vec{M_\infty}
+i k_\infty \varphi^i \vec{M_\infty}\right)\cdot\vec{n} & = \DtN_{\infty}({\varphi^i}), &&\textnormal{~on }\Gamma_\infty,\\
\vec{\nabla}\varphi^i\cdot\vec{n}&=0,&&\textnormal{~on }\Gamma,\\
\varphi^i&=g,&&\textnormal{~on }\Gamma_M.
\end{aligned}
\right.
\label{eq:pb_no_modal4}
\end{equation}

\begin{figure}[htbp]
 \centering
 \includegraphics[width=0.6\textwidth]{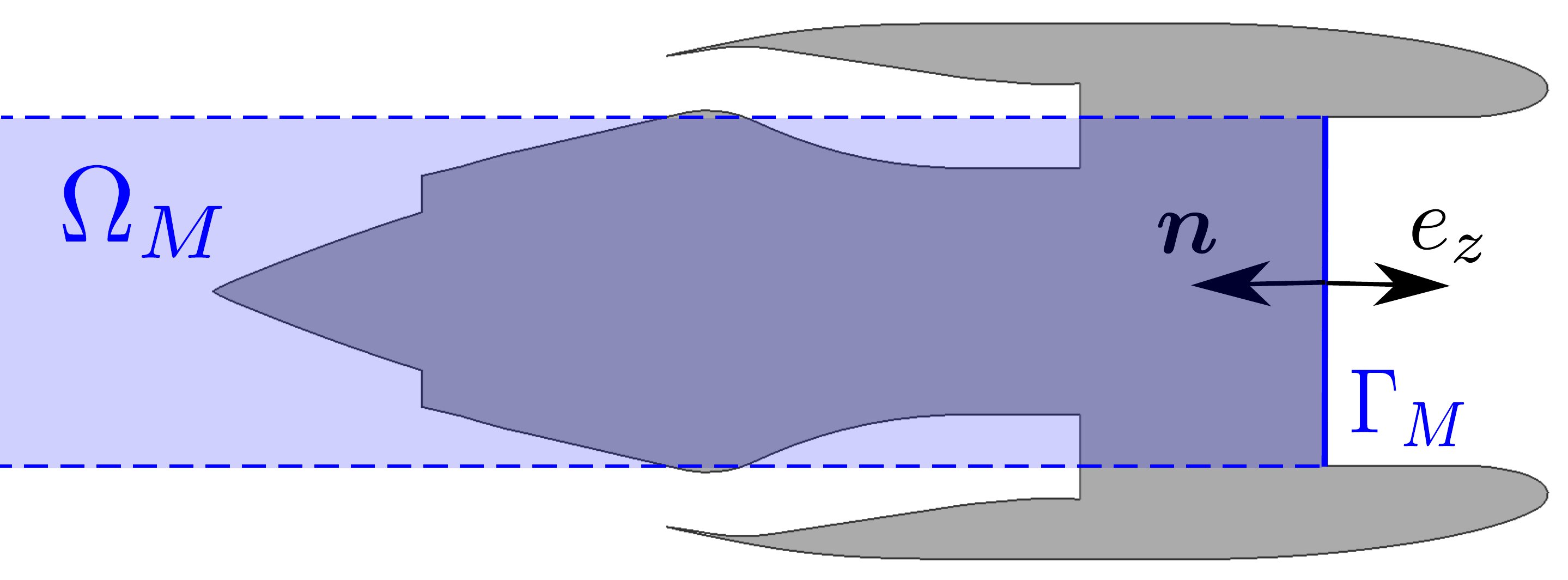}
 \caption{Model of the engine : a semi-infinite waveguide.}
 \label{fig:defomegaM}
\end{figure}

The same reasoning is done on~$\Gamma_M$, the following transmission condition holds:
$\left[\varphi\right]_{\Gamma_M}=0$ and $\left[\left(\vec{\nabla}{\varphi}
-\left(\vec{M_M}\cdot\vec{\nabla}{\varphi}\right)\vec{M_M}+i k_M \varphi \vec{M_M}\right)\cdot\vec{n}\right]_{\Gamma_M}=0$.
Then, we write a coupled problem between $\Omega^i$ and $\Omega^M$, where $\Omega^M$ is a semi-infinite waveguide with base
$\Gamma_M$ and oriented along the axis of the engine, see Figure \ref{fig:defomegaM}.
The acoustic problem in $\Omega_M$ with a Dirichlet boundary
condition on $\Gamma_M$ is well-posed, so that a Dirichlet-to-Neumann operator associated with
the problem in $\Omega_M$ can be defined, and is denoted by $\DtN_{M}$. Hence, using the transmission conditions on $\Gamma_M$,
the boundary condition at $\Gamma_M$ of the interior
problem is $\left(\vec{\nabla}{\varphi^i}-\left(\vec{M_M}\cdot\vec{\nabla}{\varphi^i}\right)\vec{M_M}
+i k_M \varphi^i \vec{M_M}\right)\cdot\vec{n}|_{\Gamma_M}=\DtN_{M}({\varphi^i}|_{\Gamma_M})$.

\subsubsection{Weak formulation}

Finally, the complete coupled problem is written
\begin{equation}
\left\{
\begin{aligned}
\mathcal{H}(\varphi^i) &=0, &&\textnormal{~in }\Omega^i,\\
\vec{\nabla}\varphi^i\cdot\vec{n}&=0,&&\textnormal{~on }\Gamma,\\
\left(\vec{\nabla}{\varphi^i}-\left(\vec{M_\infty}\cdot\vec{\nabla}{\varphi^i}\right)\vec{M_\infty}
+i k_\infty \varphi^i \vec{M_\infty}\right)\cdot\vec{n} & = \DtN_{\infty}({\varphi^i}), &&\textnormal{~on }\Gamma_\infty,\\
\left(\vec{\nabla}{\varphi^i}-\left(\vec{M_M}\cdot\vec{\nabla}{\varphi^i}\right)\vec{M_M}
+i k_M \varphi^i  \vec{M_M}\right)\cdot\vec{n} & = \DtN_{M}({\varphi^i }), &&\text{~on } \Gamma_M.
\end{aligned}
\right.
\label{eq:pb_sep}
\end{equation}
The weak formulation of the coupled problem is:
Find $\varphi^i\in H^1\left(\Omega^i\right)$ such that $\forall\varphi^t\in H^1\left(\Omega^i\right)$,
\begin{equation}
 a(\varphi^i,\varphi^t) +  I_\infty (\varphi^i,\varphi^t) + I_M (\varphi^i,\varphi^t) = 0,
\label{eq:varform_intdomain_notransf}
\end{equation}
with
\begin{align}
\begin{split}
 a(\varphi^i,\varphi^t) &=\int_{\Omega^i}\frac{\rho_0}{\rho_\infty}\biggl[ \vec{\nabla}{\varphi^i}\cdot\vec{\nabla}\bar{\varphi^t} - k_0^2  {\varphi^i}\bar{\varphi^t} \\ 
&\qquad\qquad\qquad- i k_0 \left( \left(\vec{M_0}\cdot\vec{\nabla}{\varphi^i}\right)\bar{\varphi^t}  -
\left(\vec{M_0}\cdot\vec{\nabla}{\bar{\varphi^t}}\right){\varphi^i} \right) - \left(\vec{M_0}\cdot\vec{\nabla}{\varphi^i}\right)\left(\vec{M_0}\cdot\vec{\nabla}{\bar{\varphi^t}}\right) \biggr],
\end{split}
\label{eq:form_var_volume}
\\
 I_\infty (\varphi^i,\varphi^t)  & = -\int_{\Gamma_{\infty}}
                {\DtN_{\infty}({\varphi^i})}~\overline{\varphi^t} 
    \label{eq:form_var_intinfini},\\
 I_M (\varphi^i,\varphi^t)  & = - \frac{\rho_M}{\rho_\infty}\int_{\Gamma_{M}}
                {\DtN_{M}({\varphi^i})} ~\overline{\varphi^t} 
    \label{eq:form_var_intmodal}.
\end{align}

The $\DtN_{\infty}$ map can be expressed by means of boundary integral operators written on $\Gamma_\infty$. The problem in the
exterior domain depends on $\vec{M_\infty}$. The next step is to transform \eqref{eq:varform_intdomain_notransf} in such
a way that the problem in the exterior domain becomes the classical Helmholtz solution.
This is of great interest, since we already dispose of a code evaluating the
classical boundary integral operators for the classical Helmholtz equation.

\section{Transformation of the coupled problem}
\label{sec:lorentz}

The purpose of this section is to apply a transformation to the weak formulation \eqref{eq:varform_intdomain_notransf},
such that the $\DtN_{\infty}$ map can be expressed in a convenient way. The expression of $\DtN_{M}$ resulting of the
transformation is also given, to obtained a complete transformed coupled formulation written in the same system of
coordinates.

\subsection{\PGtransf}

When the carrier flow is at rest, the acoustic potential $\varphi^e$ is solution of the Helmholtz equation.
The Prandtl-Glauert transformation was introduced by Glauert in 1928 in \cite{Glauert}, to study the
compressible effects of the air on the lift of an airfoil. This transformation was applied for subsonic aeroacoustics problems by Amiet
and Sears in \cite{Amiet} in 1970, by Astley and Bain \cite{Astley1986445} and more recently by our team in 2002 in \cite{DDMT}.

\begin{definition}
\label{prgl}
The \PGtransf associated to $\vec{M_\infty}$ consists in changing the space and time variables:
\begin{equation}
\left\{
 \begin{aligned}
  \trinf{\vec{x}}&=\vec{x}+C_\infty\left(\vec{M_\infty}\cdot\vec{x}\right)\vec{M_\infty},\\
  \trinf{t}&=t+\frac{\gamma_\infty^2}{c_\infty}\vec{M_\infty}\cdot\vec{x},
 \end{aligned}
\right.
\label{eq:def_PGtransformation}
\end{equation}
where $C_\infty=\frac{1}{M_{\infty}^2}\left(\gamma_\infty - 1 \right)$, with $\gamma_\infty=\frac{1}{\sqrt{1-M_{\infty}^2}}$.
\end{definition}

We define the spatial transformation $L$:
\begin{equation}
\mathbb{R}^3\ni\vec{x}\rightarrow L(\vec{x}):=\trinf{\vec{x}}\in\mathbb{R}^3.
\label{eq:PG_to_pb_int}
\end{equation}
The transformation $L_{|\Omega^i}$ is a dilatation of magnitude $\gamma_\infty$ along $\vec{M_\infty}$.
Denote $J_\infty$ the jacobian of $L^{-1}_{|\Omega^i}$, the inverse of this transformation. 
Consider an orthonormal basis $\left(\vec{x}', \vec{y}', \vec{z}'\right)$ of $\mathbb{R}^3$ so that $\vec{z}'=\frac{\vec{M_\infty}}{M_\infty}$.
In this basis, the jacobian matrix of $L^{-1}_{|\Omega^i}$ is 
\begin{equation}
\begin{pmatrix}
1&0&0\\
0&1&0\\
0&0&\gamma_\infty^{-1}
\end{pmatrix},
\end{equation}
so that $J_\infty=\gamma_\infty^{-1}$.

From \cite[Section 10.3.1]{wright}, applying the \PGtransf, the transformed acoustic potential in the exterior domain $f^e$
satisfies
\begin{equation}
\begin{aligned}
\trinf{\Delta} f^e + \trinf{k_{\infty}}^2 f^e = 0, &&&\textnormal{~in }\trinf{\Omega^{e}},
\end{aligned}
\label{eq:pb_sep_exterior_PGtransformed}
\end{equation}
where $\trinf{k_{\infty}}:=\gamma_\infty k_{\infty}$ is the modified wavenumber.
This is a classical Helmholtz equation. The Sommerfeld radiation condition 
\begin{equation}
\trinf{r}\left( \frac{\partial{f^e}}{\partial{\trinf{r}}}-i \trinf{k_{\infty}} f^e \right)\rightarrow 0, \qquad\trinf{r}\rightarrow +\infty
\end{equation}
is enforced as well to ensure uniqueness of the solution \cite{So12}.

\begin{remark}[Coordinate transformations]
Other coordinate transformation can retrieve the classical Helmholtz equation from the uniformly convected Helmholtz equation.
For instance, Lorentz-like transformation are possible, but would lead to frequency dependent meshes, which is not desirable.
The coordinate transformation~\eqref{eq:def_PGtransformation} we used belongs to a general family of transformations proposed in~\cite{CHAPMAN}. 
With~\eqref{eq:def_PGtransformation}, the flux of the Poynting vector is conserved through any surface orthogonal to $\vec{M}_\infty$.
\end{remark}

\begin{remark}
We define a \PGtransf associated to another vectors $\vec{v}$ by changing $\vec{M_\infty}$ and ${M_\infty}$ by respectively
$\vec{v}$ and $\|\vec{v}\|$ in Definition \ref{prgl}.
In what follows, we note by \textbf{$\trinf{\cdot}$} the objects and operators transformed by the \PGtransf associated to $\vec{M_\infty}$ (normals, geometry, derivatives).
\end{remark}

\subsection{Transformation of the volume integral term $a$ \eqref{eq:form_var_volume}}
\label{transfocoupled}

In what follows, the Prandlt-Glauert transformation is carried-out in the interior domain after the weak formulation
has been written.
Doing so, the obtained formulation is written in a form that uses operators that are already implemented
in the code \actipole \cite{justif_actipole_2011, delnevo}. Another choice can be to carry out the Prandlt-Glauert transformation
first, and then write the weak formulation. The two formulations are equivalent, but this last case suits well the
study of the existence and uniqueness of the formulation (see \cite{casenave}).

Consider the solution function $\phi^i(t,\vec{x})=\varphi^i(\vec{x})e^{-i\omega t}$. Applying the \PGtransf~\eqref{eq:def_PGtransformation}, this function transforms into
$\varphi^i(L^{-1}(\trinf{\vec{x}}))e^{ik_\infty\gamma_\infty^2\vec{M}_\infty\cdot L^{-1}(\trinf{\vec{x}})}e^{-i\omega \trinf{t}}$.
This motivates the introduction of the function $f^i$, such that $f^i(\vec{x})=\varphi^i(\vec{x})\mathcal{E}_\infty(\vec{x})$, where
$\mathcal{E}_\infty(\vec{x})=e^{ik_\infty\gamma_\infty^2\vec{M}_\infty\cdot \vec{x}}$.
We define $f^t$ from the test function $\varphi^t$ in the same fashion.
This leads to
\begin{equation}
\begin{aligned}
{\varphi^i}(\vec{x})&={f^i}(\vec{x})\overline{\mathcal{E}_\infty}(\vec{x})\\
\overline{\varphi^t}(\vec{x})&=\overline{f^t}(\vec{x}){\mathcal{E}_\infty}(\vec{x}).
\label{eq:zeroorder}
\end{aligned}
\end{equation}
The transformation $H^1(\Omega^i)\ni\varphi^t({\vec{x}})\rightarrow f^t({\vec{x}}) {\mathcal{E}_\infty}(\vec{x})\in
H^1(\trinf{\Omega^i})$ is surjective, and therefore this modification of test function is still compatible with the weak formulation.
The gradients are transformed to
\begin{equation}
\begin{aligned}
\vec{\nabla}{\varphi^i}(\vec{x})&=\left(\vec{\nabla}{f^i}(\vec{x})-ik_\infty\gamma_\infty^2\vec{M}_\infty{f^i}(\vec{x})\right)\overline{\mathcal{E}_\infty}(\vec{x})\\
\vec{\nabla}\overline{\varphi^t}(\vec{x})&=\left(\vec{\nabla}\overline{f^t}(\vec{x})+ik_\infty\gamma_\infty^2\vec{M}_\infty\overline{f^t}(\vec{x})\right){\mathcal{E}_\infty}(\vec{x}).
\label{eq:oneorder}
\end{aligned}
\end{equation}
Thus, Equation~\eqref{eq:form_var_volume} becomes
\begin{equation}
\begin{aligned}
 a(\varphi^i,\varphi^t) &= 
  \int_{{\Omega^i}}\frac{\rho_0}{\rho_\infty}\left[ \left({\vec{\mathcal{L_-}}}{f^i}\right)\cdot\left({\vec{\mathcal{L_+}}}\bar{f^t}\right) - k_0^2  {f^i} \bar{f^t} - i {k_0} \left( \left( \vec{{M_0}}\cdot\left({\vec{\mathcal{L_-}}}{f^i}\right)\right) \bar{f^t} \right.\right.\\
&\qquad\left.\left. - \left( \vec{{M_0}}\cdot\left({\vec{\mathcal{L_+}}}\bar{f^t}\right) \right) {f^i} \right)\right]- \int_{{\Omega^i}}\frac{\rho_0}{\rho_\infty}\left( \vec{{M_0}}\cdot\left({\vec{\mathcal{L_-}}}{f^i}\right) \right)\left( \vec{{M_0}}\cdot\left({\vec{\mathcal{L_+}}}\bar{f^t}\right) \right),
\label{eq:define_variational_volume_operator0}
\end{aligned}
\end{equation}
where ${\vec{\mathcal{L_{\pm}}}}:={\vec\nabla} \pm i k_{\infty}\gamma_\infty^2\vec{M_{\infty}}$
and where we used $\bar{\mathcal{E}_\infty}({\vec{x}}){\mathcal{E}_\infty}({\vec{x}})=1$.

Then, applying the change of variables and making use of the jacobian $J_\infty$ of $L^{-1}_{|\Omega^i}$, there holds
\begin{equation}
\begin{aligned}
 a(\varphi^i,\varphi^t) &= 
 J_\infty \int_{\trinf{\Omega^i}}\frac{\rho_0}{\rho_\infty}\left[ \left(\trinf{\vec{\mathcal{L_-}}}{f^i}\right)\cdot\left(\trinf{\vec{\mathcal{L_+}}}\bar{f^t}\right) - k_0^2  {f^i} \bar{f^t} - i {k_0} \left( \left( \vec{{M_0}}\cdot\left(\trinf{\vec{\mathcal{L_-}}}{f^i}\right)\right) \bar{f^t} \right.\right.\\
&\qquad\left.\left. - \left( \vec{{M_0}}\cdot\left(\trinf{\vec{\mathcal{L_+}}}\bar{f^t}\right) \right) {f^i} \right)\right]- J_\infty \int_{\trinf{\Omega^i}}\frac{\rho_0}{\rho_\infty}\left( \vec{{M_0}}\cdot\left(\trinf{\vec{\mathcal{L_-}}}{f^i}\right) \right)\left( \vec{{M_0}}\cdot\left(\trinf{\vec{\mathcal{L_+}}}\bar{f^t}\right) \right),
\label{eq:define_variational_volume_operator}
\end{aligned}
\end{equation}
where $\trinf{\vec{\mathcal{L_{\pm}}}}:=\trinf{\vec\nabla} +
\left(C_\infty \vec{M_{\infty}}\cdot\trinf{\vec\nabla} \pm i k_{\infty}\gamma_\infty^2\right)\vec{M_{\infty}}$.
Notice that when changing the variables, we wrote the differential operators in the transformed system of coordination, using
\begin{equation}
\vec{\nabla}\rightarrow \trinf{\vec{\nabla}}+C_\infty\vec{M_\infty} \, \vec{M_\infty}\cdot\trinf{\vec{\nabla}},
\label{eq:transfgrad}
\end{equation}
which is directly derived from the first line of~\eqref{eq:def_PGtransformation}.

\begin{remark}
If we impose $\vec{M}_0=\vec{M}_\infty$, $\rho_0=\rho_\infty$ and $c_0=c_\infty$ in~\eqref{eq:define_variational_volume_operator}, we find
\begin{equation}
 a(\varphi^i,\varphi^t) =  J_\infty \int_{\trinf{\Omega^i}}
\left(\trinf{\vec{\nabla}}{f^i}\cdot\trinf{\vec{\nabla}}\bar{f^t}- \trinf{k_\infty}^2  {f^i} \bar{f^t}\right),
\end{equation}
which is the variational formulation for the nonconvected Helmholtz equation with modified wavenumber $\trinf{k_\infty} = \gamma_\infty k_\infty$ on $f$ on the transformed geometry
and in the transformed coordinates.
\end{remark}

\begin{remark}
The expression~\eqref{eq:define_variational_volume_operator} for the volume integral is more complicated than the 
expression~\eqref{eq:form_var_volume}. However, this will enable us to treat the coupling with the exterior domain in a simple way.
\end{remark}

\subsection{Transformation of the surface integral term $I_\infty$ \eqref{eq:form_var_intinfini}}
\label{sec:transfointeq}

\subsubsection{Transformation of normals}

Consider a closed surface $\mathcal{C}$ defined by a function $\Phi$ by $\Phi(\vec{x})=0$, and denote by $\trinf{\mathcal{C}}$ the transformation of $\mathcal{C}$ by $L$.
Let $\vec{x}\in\mathcal{C}$, the normal to $\mathcal{C}$ at $\vec{x}$, $\vec{n}(\vec{x})$, is colinear to $\nabla{\Phi}(\vec{x})$, whereas the normal to $\trinf{\mathcal{C}}$
at $\trinf{\vec{x}}$, $\trinf{\vec{n}}(\trinf{\vec{x}})$, is colinear to $\trinf\nabla{\Phi}(\trinf{\vec{x}})$.
From~\eqref{eq:transfgrad}, there then holds
\begin{equation}
\vec{n}(\vec{x})=K_\infty(\trinf{\vec{x}})\left(\trinf{\vec{n}}\left(\trinf{\vec{x}}\right) + C_\infty \left(\vec{M_\infty}\cdot
\trinf{\vec{n}}\left(\trinf{\vec{x}}\right)\right)\vec{M_\infty}\right),
\label{eq:expK}
\end{equation}
where $K_\infty(\trinf{\vec{x}})$ is a normalization factor.
Taking the square of the norm of both sides of~\eqref{eq:expK}, there holds $1=K_\infty(\trinf{\vec{x}})^2\left(1+
(M_\infty^2C_\infty^2+2C_\infty)\left(\vec{M_\infty}\cdot\trinf{\vec{n}}\left(\trinf{\vec{x}}\right)\right)^2\right)$, from which we deduce
\begin{equation}
 K_\infty(\trinf{\vec{x}})=\sqrt{1+\left(\gamma_\infty\vec{M_\infty}\cdot\trinf{\vec{n}}\left(\trinf{\vec{x}}\right)\right)^2}.
\label{eq:K_expression}
\end{equation}

Conversely, to express $\trinf{\vec{n}}\left(\trinf{\vec{x}}\right)$ as a function of $\vec{n}(\vec{x})$, we consider the inverse spatial transformation $L^{-1}$,
which consists of a contraction of magnitude $\gamma_\infty^{-1}$ along $\vec{M}_\infty$. The gradient are then changed as
$\trinf{\vec{\nabla}}\rightarrow \vec{\nabla}+\trinf{C_\infty}\vec{M_\infty} \, \vec{M_\infty}\cdot\vec{\nabla}$,
where $\trinf{C_\infty}=\frac{1}{M_{\infty}^2}\left( \gamma_\infty^{-1} - 1 \right)$. Applying this to the gradient of $\Phi(\trinf{x})=0$, which defines
 $\trinf{\mathcal{C}}$, we deduce
\begin{equation}
\trinf{\vec{n}}(\trinf{\vec{x}})=\trinf{K_\infty}(\vec{x})\left(\vec{n}\left(\vec{x}\right) + \trinf{C_\infty} \left(\vec{M_\infty}\cdot
\vec{n}\left(\vec{x}\right)\right)\vec{M_\infty}\right),
\label{eq:nprime_fonction_n}
\end{equation}
where $\trinf{K_\infty}(\vec{x})$ is a normalization factor.
Using the normalization condition,
\begin{equation}
 \trinf{K_\infty}(\vec{x})=\frac{1}{\sqrt{1-\left(\vec{M_\infty}\cdot\vec{n}\left(\vec{x}\right)\right)^2}}.
\label{eq:Kprime_expression}
\end{equation}

\subsubsection{Jacobian of the spatial transformation restricted to $\Gamma_\infty$}

\begin{figure}[h!]
 \centering
\includegraphics[width=0.6\textwidth]{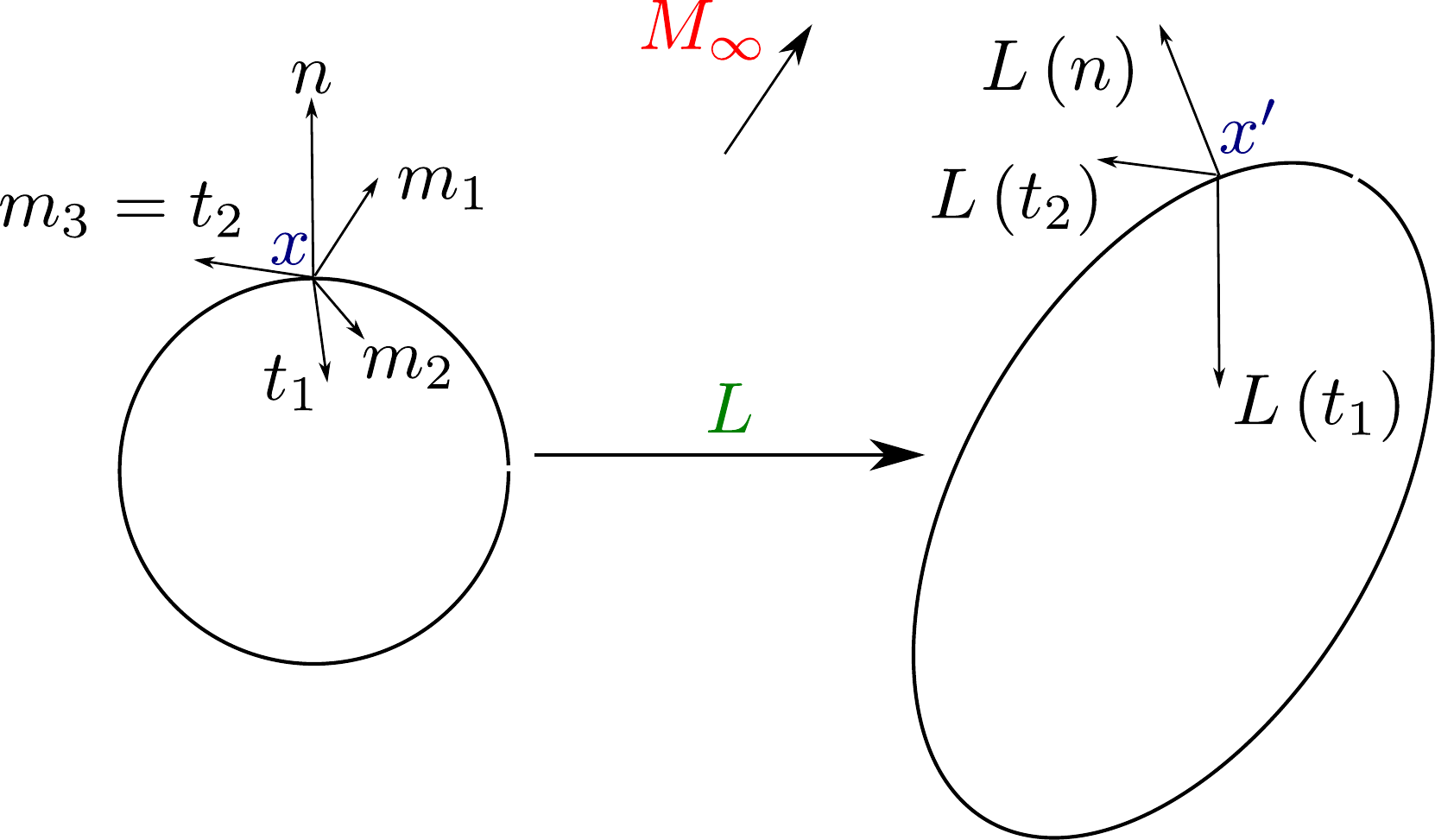}
 \caption{Transformation of a local basis on $\Gamma_\infty$ by the \PGtransf.}
\label{fig:figproof}
\end{figure}

Consider $L_{|\Gamma_\infty}$, the restriction of $L$ to the $2$-dimensional manifold $\Gamma_\infty$.
Let $\vec{x}\in\Gamma_\infty$. Let $\vec{m_1}:=\frac{\vec{M_\infty}}{M_\infty}$,
$\vec{m_2}:=\frac{\vec{n}-(\vec{m_1}\cdot\vec{n})\vec{m_1}}{\|\vec{n}-(\vec{m_1}\cdot\vec{n})\vec{m_1}\|}$ and $\vec{m_3}$ such that $\left(\vec{m_1}, \vec{m_2}, \vec{m_3}\right)$ is
an orthonormal triplet. Let $N:=\vec{n}\cdot\vec{m_1}$. By construction, $\vec{n}=N\vec{m_1}+\sqrt{1-N^2}\vec{m_2}$.
Then, we define $\vec{t_1}:=-\sqrt{1-N^2}\vec{m_1}+N\vec{m_2}$ and $\vec{t_2}:=\vec{m_3}$, see Figure \ref{fig:figproof}.
It is direct to verify that $\left(\vec{t_1}, \vec{t_2}\right)$ is an orthonormal doublet,
and that $\vec{t_1}\cdot\vec{n}=\vec{t_2}\cdot\vec{n}=0$. Therefore, $\left(\vec{t_1},\vec{t_2}\right)$ is an orthonormal basis for the
hyperplane tangent to $\Gamma_\infty$ in $\vec{x}$, and $|Jac(L_{|\Gamma_\infty})|=|\det\left(L(\vec{t_1}),L(\vec{t_2})\right)|$.
Since we have identified in $\vec{t_1}$ and $\vec{t_2}$ the components parallel and orthogonal to $\vec{M_\infty}$, there holds
$L(\vec{t_1})=-\gamma_\infty\sqrt{1-N^2}\vec{m_1}+N\vec{m_2}$ and $L(\vec{t_2})=\vec{t_2}=\vec{m_3}$, because $\vec{t_2}$ is orthogonal to $\vec{M}_\infty$.
We see that $L(\vec{t_1})$ and $L(\vec{t_2})$ are orthogonal, then $\det\left(L(\vec{t_1}),L(\vec{t_2})\right) = \|L(\vec{t_1})\|\|L(\vec{t_2})\| = \|L(\vec{t_1})\| =
\sqrt{N^2+\gamma_\infty^2(1-N^2)}=\gamma_\infty\sqrt{1-(\vec{M_\infty}\cdot\vec{n})^2}$, from which we deduce
$|Jac(L^{-1}_{|\Gamma_\infty})|=J_\infty \trinf{K_\infty}$.

\subsubsection{Expression for $I_\infty$}

Consider equation~\eqref{eq:form_var_intinfini} for $I_\infty$ and use the boundary condition line~3 of~\eqref{eq:pb_sep}:
\begin{equation}
 I_\infty (\varphi^i,\varphi^t)  = -\int_{\Gamma_{\infty}}{\left(\vec{\nabla}{\varphi^i}-\left(\vec{M_\infty}\cdot\vec{\nabla}{\varphi^i}\right)\vec{M_\infty}
-i k_\infty {\varphi^i} \vec{M_\infty}\right)\cdot\vec{n}}~{\varphi^t} .
\end{equation}
Plugging the changes for order zero~\eqref{eq:zeroorder} and order one~\eqref{eq:oneorder} terms and making use of the jacobian $Jac(L^{-1}_{|\Gamma_\infty})$, there holds
\begin{multline}
 I_\infty (\varphi^i,\varphi^t) = -J_\infty \int_{\trinf{\Gamma_\infty}}\trinf{K_\infty}\Biggl[ \left(\trinf{\vec\nabla}{f^i}-ik_\infty\gamma_\infty\vec{M_\infty}{f^i}\right) \\
\cdot\left(\vec{{n}}+\left(C_\infty-\gamma_\infty\right)\left(\vec M_{\infty}\cdot\vec{{n}}\right)\vec M_{\infty}\right) 
 +i k_{\infty}{f^i}\left(\vec{M_{\infty}}\cdot\vec{{n}}\right)   \Biggr] \bar{f^t}.
\end{multline}
Since $C_\infty-\gamma_\infty=\trinf{C_\infty}$, we recognize the expression of $\frac{\trinf{\vec{n}}}{\trinf{K_\infty}}$. Reorganizing the terms:
\begin{equation}
 I_\infty (\varphi^i,\varphi^t)=-J_\infty \int_{\trinf{\Gamma_\infty}}{\trinf{\vec\nabla}{f^i}\cdot\trinf{\vec{n}}}~\bar{f^t}+J_\infty \int_{\trinf{\Gamma_\infty}}{ik_\infty {f^i}\left(
\trinf{K_\infty}\left(\vec{M_\infty}\cdot\vec{n}\right)-\gamma_\infty\vec{M_\infty}\cdot\trinf{\vec{n}}\right)\bar{f^t}}.
\end{equation}
Finally, from relation \eqref{eq:nprime_fonction_n}, $\vec{M_\infty}\cdot\trinf{\vec{n}}=\frac{\trinf{K_\infty}}{\gamma_\infty}\left(\vec{M_\infty}\cdot\vec{n}\right)$,
and
\begin{equation}
 I_\infty (\varphi^i,\varphi^t)=-J_\infty \int_{\trinf{\Gamma_\infty}}{ \trinf{\vec\nabla} {f^i}\cdot\trinf{\vec{n}}~\bar{f^t}}.
\label{eq:I_infty}
\end{equation}

\begin{remark}
Notice the extreme simplification of the surface integral term \eqref{eq:I_infty}.
The direct coupling with the BEM is possible thanks to this particular form of the surface integral term.
Notice also that~\eqref{eq:I_infty} is simply the expression of the surface integral term of the weak formulation~\eqref{eq:varform_intdomain_notransf} after the \PGtransf has been applied.
\end{remark}

We now need to apply the \PGtransf in the equation on the exterior domain~$\Omega^e$ and we will see that~\eqref{eq:I_infty} can be expressed using the
Dirichlet-to-Neumann operator associated to the classical Helmholtz equation.

\subsection{Derivation of the transformed $\trinf{\DtN_{\infty}}$ map}

The derivation of a Dirichlet-to-Neumann map $\DtN_\infty$ for the Helmholtz exterior problem \eqref{eq:pb_sep_exterior_PGtransformed} is
classical. The procedure is detailed for instance in \cite{casenave}. We recall the main steps for completeness of the
presentation.
A function is called a radiating Helmholtz solution, if it solves the classical Helmholtz solution in $\Omega^e$ and in
$\mathbb{R}^3\backslash\overline{\Omega^e}$, and if it satisfies the Sommerfeld radiation condition.
A radiating Helmholtz solution $v$ can be represented from the jump of its traces on $\Gamma$ \cite[Theorem 3.1.1]{Ne01}
\begin{equation}
\label{eq:repfrom}
v = -\mathcal{S}\left[\vec{\nabla}v\cdot\vec{n}\right] + \mathcal{D}\left[v\right]\textnormal{~~~in }\Omega^{e}\cup\mathbb{R}^3\backslash\Omega^e,
\end{equation} 
where $\mathcal{S}$ and $\mathcal{D}$ are the single-layer and double-layer potentials associated with the Helmholtz equation, defined such that
\begin{align*}
 \mathcal{S} \lambda (\vec{x}) & =  \int_{\Gamma_\infty} G(\vec{x},\vec{y}) \lambda(\vec{y}) \, d\vec{y} , &
 \mathcal{D} \mu  (\vec{x}) & =     \int_{\Gamma_\infty}\frac{\partial G(\vec{x},\vec{y})}{\partial n_x}\lambda( \vec{y}) \, d\vec{y},
& \forall \vec{x}\in \Omega^{e}\cup\mathbb{R}^3\backslash\Omega^e,
\end{align*}
where $G(\vec{x},\vec{y}):=\frac{\exp\left(-ik_\infty\left|\vec{x}-\vec{y}\right|\right)}{4\pi\left|\vec{x}-\vec{y}\right|}$
is the fundamental solution to the classical Helmholtz equation with wavenumber $k_\infty$.
If $v$ is a radiating Helmholtz solution, then \cite[Theorem 3.1.2]{Ne01}
\begin{equation}
\label{eq:caldproj}
\left[
\begin{array}{cc}
 \frac{1}{2}I - D & S\\
 N & \frac{1}{2}I + D^*
\end{array}
\right]\left[
\begin{array}{cc}
\left[v\right]_{\Gamma_\infty}\\
\left[\vec{\nabla}v\cdot\vec{n}\right]_{\Gamma_\infty}
\end{array}\right]=-\left[
\begin{array}{cc}
v^{i}\\
\vec{\nabla}v^{i}\cdot\vec{n}
\end{array}
\right]\textnormal{~~~on }\Gamma_\infty,
\end{equation}
where $S$, $D$, $D^*$ and $N$ are respectively the single-layer, double-layer, transpose of the double-layer and hypersingular
boundary integral operators defined as, $\forall \vec{x} \in \Gamma_\infty$,
\begin{align*}
S \lambda(\vec{x})&=\int_{\Gamma_\infty} G(\vec{x},\vec{y}) \lambda(\vec{y}) d\vec{y}, &
D \mu(\vec{x})&=\int_{\Gamma_\infty} \frac{\partial G(\vec{x},\vec{y})}{\partial n_y} \mu(\vec{y}) d\vec{y},\\
D^* \lambda(\vec{x})&=\int_{\Gamma_\infty} \frac{\partial G(\vec{x},\vec{y})}{\partial n_x}\lambda(\vec{y}) d\vec{y}, &
N \mu(\vec{x})&=\oint_{\Gamma_\infty} \frac{\partial^2 G(\vec{x},\vec{y})}{\partial n_x \partial n_y} \mu(\vec{y}) d\vec{y}.
\end{align*}

\noindent
Consider the function $u$ such that $u|_{\trinf{\Omega^e}}:=f^e$ and $u|_{\trinf{\Omega^i}}:=0$, where $f^e$ solves \eqref{eq:pb_sep_exterior_PGtransformed},
so that $u$ is a radiating Helmholtz solution in the transformed coordinates and on the transformed geometry. 
\begin{remark}
We chose $u|_{\trinf{\Omega^i}}:=0$ to ensure that $u|_{\trinf{\Omega^i}}$ solves the Helmholtz equation in $\Omega^i$, which is clearly the case, and to ensure
that $u$ is a radiating Helmholtz solution, to apply~\eqref{eq:caldproj}. We could have chosen any solution to the Helmholtz equation in $\Omega^i$ for $u|_{\trinf{\Omega^i}}$,
but in this case, the relation~\eqref{eq:caldproj} would have involve jumps of traces of some function not related to the problem.
Our choice is motivated by the fact that, with $u|_{\trinf{\Omega^e}}:=f^e$ and $u|_{\trinf{\Omega^i}}:=0$, $\left[u\right]_{\trinf{\Gamma_\infty}}=f^e|_{\trinf{\Gamma_\infty}}$ and
$\left[\vec{\nabla}u\cdot\vec{n}\right]_{\trinf{\Gamma_\infty}}=\trinf{\vec{\nabla}}f^e\cdot\trinf{\vec{n}}|_{\trinf{\Gamma_\infty}}$, so that the relation~\eqref{eq:caldproj}
only involves the total transformed acoustic potential on $\trinf{\Gamma_\infty}$, which will provide a simple coupling with~\eqref{eq:I_infty}.
\end{remark}

Consider the transmission conditions~\eqref{eq:pb_no_modal2}, lines 3 and 4. Applying the \PGtransf (using the first line of~\eqref{eq:zeroorder} and~\eqref{eq:oneorder}) and
making use of~\eqref{eq:expK}, there holds
\begin{equation}
\left\{
\begin{aligned}
\left[f\right]_{\trinf{\Gamma_\infty}}&=0,\\
\left[\trinf{\vec{\nabla}}f\cdot\trinf{\vec{n}}\right]_{\trinf{\Gamma_\infty}}&=0,
\end{aligned}
\right.
\end{equation}
which we can write
\begin{equation}
\begin{aligned}
f^i&=f^e=f,&\textnormal{~~~on }\trinf{\Gamma_\infty},\\
\trinf{\vec{\nabla}}f^i\cdot\trinf{\vec{n}}&=\trinf{\vec{\nabla}}f^e\cdot\trinf{\vec{n}}=\trinf{\vec{\nabla}}f\cdot\trinf{\vec{n}},&\textnormal{~~~on }\trinf{\Gamma_\infty}.
\end{aligned}
\end{equation}
Using \eqref{eq:caldproj} with $v=u$ yields

\begin{equation}
\label{eq:DtNsys}
\left[
\begin{array}{cc}
 \frac{1}{2}I - \trinf{D} & \trinf{S}\\
 \trinf{N} & \frac{1}{2}I + \trinf{{D^*}}
\end{array}
\right]\left[
\begin{array}{cc}
f\\
\trinf{\vec{\nabla}}f\cdot\trinf{\vec{n}}
\end{array}\right]=\left[
\begin{array}{cc}
0\\
0
\end{array}
\right],\textnormal{~~~on }\trinf{\Gamma_\infty},
\end{equation}
where $\trinf{S}$, $\trinf{D}$, $\trinf{D^{*}}$ and $\trinf{N}$ are respectively the single-layer, double-layer, transpose of the double-layer and hypersingular
boundary integral operators for the Helmholtz equation obtained in the Prandtl--Glauert transformed space with modified
wavenumber $\trinf{k_\infty}$:
\begin{align*}
\trinf{S} \lambda  (\trinf{\vec{x}}) &=\int_{\trinf{\Gamma_\infty}} \hat{G}(\trinf{\vec{x}},\trinf{\vec{y}}) \lambda(\trinf{\vec{y}}) d\trinf{\vec{y}}, &
\trinf{D} \mu      (\trinf{\vec{x}}) &=\int_{\trinf{\Gamma_\infty}} \frac{\partial \hat{G}(\trinf{\vec{x}},\trinf{\vec{y}})}{\partial \trinf{n_y}} \mu(\trinf{\vec{y}}) d\trinf{\vec{y}},\\
\trinf{D^*} \lambda(\trinf{\vec{x}}) &=\int_{\trinf{\Gamma_\infty}} \frac{\partial \hat{G}(\trinf{\vec{x}},\trinf{\vec{y}})}{\partial \trinf{n_x}} \lambda(\trinf{\vec{y}}) d\trinf{\vec{y}}, &
\trinf{N} \mu      (\trinf{\vec{x}}) &=\oint_{\trinf{\Gamma_\infty}} \frac{\partial^2 \hat{G}(\trinf{\vec{x}},\trinf{\vec{y}})}{\partial \trinf{n_x} \partial \trinf{n_y}} \mu(\trinf{\vec{y}}) d\trinf{\vec{y}},
\end{align*}
where $\hat{G}(\vec{x},\vec{y}):=\frac{\exp\left(-i\trinf{k_\infty}\left|\vec{x}-\vec{y}\right|\right)}{4\pi\left|\vec{x}-\vec{y}\right|}$
is the fundamental solution to the classical Helmholtz equation with modified wavenumber $\trinf{k_\infty}$.

Using the first line of~\eqref{eq:DtNsys}, there holds
\begin{equation}
\label{eq:step1defDtN}
\trinf{\vec{\nabla}}f\cdot\trinf{\vec{n}}=\trinf{S}^{-1}\left(\trinf{D}-\frac{1}{2}I\right)(f).
\end{equation}
Then, we subtract $\trinf{\vec{\nabla}}f\cdot\trinf{\vec{n}}$ from the second line of~\eqref{eq:DtNsys} and inverse the signs to obtain
\begin{equation}
\trinf{\vec{\nabla}}f\cdot\trinf{\vec{n}}=-\trinf{N}(f)+\left(\frac{1}{2}I-\trinf{D^*}\right)\left(\trinf{\vec{\nabla}}f\cdot\trinf{\vec{n}}\right).
\label{eq:comp_DtN}
\end{equation}
Injecting \eqref{eq:step1defDtN} into the right-hand side of~\eqref{eq:comp_DtN}, an expression of the
$\trinf{\DtN_{\infty}}$ map can be obtained in the following form:
\begin{equation}
\label{eq:defDtNinftrans}
\trinf{\vec{\nabla}}f\cdot\trinf{\vec{n}}=\trinf{\DtN_{\infty}}\left(f\right):=-\trinf{N}(f)+\left(\frac{1}{2}I-\trinf{D^*}\right)
\trinf{S}^{-1}\left(\trinf{D}-\frac{1}{2}I\right)(f).
\end{equation}
Notice that other $\trinf{\DtN_{\infty}}$ maps can be readily obtained from~\eqref{eq:DtNsys}. Our choice leads to a symmetric linear system (see the matrix~\eqref{eq:big_matrix}),
for which computation optimizations can be used.

\begin{remark}
Even though it is possible to write integral equations for the uniformly convected Helmholtz equation (see \cite{inteqconvHelmholtz}),
the Prandtl--Glauert transformation allows us to write integral equation that only involves the Green kernel associated to
the Helmholtz equation. Hence, we can profit from our validated code \actipole developed by our team \cite{justif_actipole_2011, delnevo}.
\end{remark}

\subsection{Computation of the coupling integral term $I_M$ \eqref{eq:form_var_intmodal}}
\label{sec:coupling_integral_term_form_classic}

The engine is modeled by a semi-infinite waveguide with the classical hypothesis that the flow is uniform. To simplify the presentation, the waveguide is also supposed to be oriented along the local axis $\vec{e_{z}}$ so that $\Gamma_M$ is orthogonal to $\vec{e_{z}}$
(see Figure \ref{fig:defomegaM}). 
More precisely, we suppose here that $\Gamma_M$ is included in the plane $z=0$. The flow is defined by $\vec{M_M}$ and is then parallel to $\vec{e_{z}}$. 
A more general formulation is presented in appendix \ref{sec:appendix_form_modes}.

In $\Omega_M$, the acoustic potential is decomposed into an incident and a diffracted potential:
$\varphi:=\varphi^{\rm inc}+\varphi^{\rm diff}$, both solution to the following convected Helmholtz equation:
\begin{equation}
\label{eq:modalpb}
\Delta\varphi^{\rm inc,diff}+k_{M}^2\varphi^{\rm inc,diff}+2ik_{M}\vec{M_M}\cdot\vec{\nabla}\varphi^{\rm inc,diff}
-\vec{M_M}\cdot\vec{\nabla}\left( \vec{M_M}\cdot\vec{\nabla}\varphi^{\rm inc,diff} \right)=0 \quad\text{in } \Omega_M.
\end{equation}
The incident potential is known, whereas the diffracted potential is unknown. 
Under these assumptions, the following decomposition holds \cite{MI68, Br06}:
\begin{equation}
\begin{aligned}
 \varphi^{\inc}(x,y,z) &=\sum_{(\mtimesn) }   
\alpha_{\mn}  \, v_{\mn}(x,y) \exp\left({ik_{\mn}^{+} z}\right)\quad\text{in } \Omega_M,\\
 \varphi^{\diffr}(x,y,z)& = \sum_{(\mtimesn) }   
\beta_{\mn} \,  v_{\mn}(x,y)  \exp\left({ik_{\mn}^{-} z}\right)\quad\text{in } \Omega_M,
\label{eq:def_decomp_modal}
\end{aligned}
\end{equation}
with $\mn$ a discrete index and where the incident modal coefficients $\alpha_{\mn} := \int_{\Gamma_M}\varphi^{\inc}\bar{v_{\mn}}$ are supposed known
(input of the problem) while the diffracted modal coefficients $\beta_{\mn} := \int_{\Gamma_M}\varphi^{\diffr}\bar{v_{\mn}}$ are some unknowns of the problem.
The basis functions $v_{\mn}$ constitute a modal basis function chosen to be orthonormal. 

For instance for a cylindrical duct of radius $R$, $\mn$ corresponds to a couple of indices $(m,n) \in (\Z\times\N^{*})$ and the functions  $v_{\mn}$ are defined in polar coordinates by \cite{MI68, Br06} 
\begin{equation}
  v_{\mn}(r,\theta) = v_{m,n}(r,\theta) :=  V_{m,n} \; J_m\left(\frac{r_{m,n}}{R} r\right) \exp\left({i m \theta}\right),
\end{equation}
with $r_{m,n}$ the $n$-th zero of the derivative of $m$-th Bessel function of the first kind $J_m$, 
$V_{m,n}$ the normalization factor such that $\int_{\Gamma_M} v_{m,n}^2 = 1$, 
and 
\begin{align}
k_\mn^{\pm}= k_{mn}^{\pm} &= \frac{-k_M M_M \pm \sqrt{k_M^2 - \left( 1-M_M^2 \right)\left(\displaystyle \frac{r_{m,n}}{R}\right)^2}}{1 - M_M^2} && \text{~ for propagating modes}  ~ (k_{mn}^{\pm} \in \mathbb{R}),\\ 
k_\mn^{\pm}= k_{mn}^{\pm} &= \frac{-k_M M_M \pm i \sqrt{\left( 1-M_M^2 \right)\left(\displaystyle \frac{r_{m,n}}{R}\right)^2 - k_M^2}}{1 - M_M^2} && \text{~ for evanescent modes}   ~ ( k_{mn}^{\pm} \in \mathbb{C}),
\end{align}
the wavenumber of each mode. 
For any $(m,n)\in(\mathbb{Z}\times\mathbb{N}^*)$, the corresponding mode is either propagating or evanescent. 
For any shape of the duct, there exist a finite number of propagating modes and an infinite number of evanescent modes.

Based on this decomposition, the expression of the Dirichlet-to-Neumann operator $\DtN_M$ is \cite{dahi_conditionD2N_modes:2002, PhDlegendre:2003}:
\begin{equation}
\left(\vec{\nabla}{\varphi}-\left(\vec{M_M}\cdot\vec{\nabla}{\varphi}\right)\vec{M_M}
+i k_M \varphi  \vec{M_M}\right)\cdot\vec{n}=
\DtN_M \left(  \varphi \right) := \sum_{(\mtimesn) } 
             \left(\alpha_{\mn}  Y^{+}_{\mn} +\beta_{\mn}  Y^{-}_{\mn}\right)v_{\mn}\quad\textnormal{on }\Gamma_M,
\label{eq:def_operator_DtN}
\end{equation}
where
\begin{align}
  Y^{\pm}_{\mn}  & := -i \left[ k_{\mn}^{\pm} \left( 1 - M_M^2 \right) + k_M M_M \right] 
\end{align}

By definition, 
\begin{align*}
 I_M (\varphi,\varphi^t)  & =
 - \frac{\rho_M}{\rho_\infty}\int_{\Gamma_{M}}
               \DtN_{M}({\varphi}) ~\overline{{\varphi^t}}  \\
& =  -  \frac{\rho_M}{\rho_\infty}  \sum_{(\mtimesn)} \alpha_{\mn} Y_{\mn}^+  \int_{{\Gamma_M}} v_{\mn} \, \overline{\varphi^t}
     -  \frac{\rho_M}{\rho_\infty}  \sum_{(\mtimesn)} \beta_{\mn} Y_{\mn}^-  \int_{{\Gamma_M}} v_{\mn} \, \overline{\varphi^t}.
\end{align*}

Notice that since $\Gamma_M$ is included in the plane $z=0$ and
$M_\infty$ is directed along $\vec{e_z}$, then ${\trinf{\Gamma_M}}=\Gamma_M$, and $\trinf{v_{\mn}}={v_{\mn}}$ on $\Gamma_M$,
where $\trinf{v_{\mn}}$ is the \PGtransf of $v_{\mn}$.
Hence, $\alpha_{\mn} = \int_{\trinf{\Gamma_M}}f^{\inc}\overline{\trinf{v_{\mn}}}$
and $\beta_{\mn} = \int_{\trinf{\Gamma_M}}f^{\diffr}\overline{\trinf{v_{\mn}}}$,
where $f^{\inc}$ and $f^{\diffr}$ are the \PGtransf of respectively $\varphi^{\inc}$ and $\varphi^{\diffr}$.
In view of the coupling in the coordinates and geometry transformed by the \PGtransf, we can write
\begin{equation}
  I_M (\varphi,\varphi^t) 
=  - \frac{\rho_M}{\rho_\infty}  \sum_{(\mtimesn)} \left(\alpha_{\mn} Y_{\mn}^+ + \beta_{\mn} Y_{\mn}^- \right)\int_{\trinf{\Gamma_M}} \trinf{v_{\mn}}\overline{f^t}
\label{eq:surf_int_mod_bonrepere_v0}
\end{equation}
with $f^t$ the \PGtransf of $\varphi^t$.
Let us define $\gamma_{\pq}$ by $\dsp f^t=\sum_{(\ptimesq)} \gamma_{\pq}\trinf{v_{\pq}}$. Then, using the orthonormality of the modal basis
$\int_{\Gamma_M}\trinf{v_{\mn}}\bar{\trinf{v_{\pq}}} =  \delta_{\mn,\pq}$, where $\delta_{\mn,\pq}$ refers to the Kronecker delta, we obtain
\begin{equation}
  I_M (\varphi,\varphi^t) 
=  - \frac{\rho_M}{\rho_\infty}  \sum_{(\mtimesn)} \left(\alpha_{\mn} Y_{\mn}^+ + \beta_{\mn} Y_{\mn}^- \right) \bar{\gamma_{\mn}}.
\label{eq:surf_int_mod_bonrepere_v01}
\end{equation}
To write a direct coupled problem, it is more practical to consider the coefficient of decomposition of the total acoustic potential.
Consider $\varsigma_{\mn} := \int_{\Gamma_M}\varphi \, \bar{v_{\mn}}=\alpha_{\mn}+\beta_{\mn}$,
there holds
\begin{equation}
  I_M (\varphi,\varphi^t) 
=  - \frac{\rho_M}{\rho_\infty}  \sum_{(\mtimesn)} \left(\alpha_{\mn} \left(Y_{\mn}^+-Y_{\mn}^-\right) + \varsigma_{\mn} Y_{\mn}^- \right)\bar{\gamma_{\mn}}.
\label{eq:surf_int_mod_bonrepere_v1}
\end{equation}

\subsection{Transformed coupled problem}

To treat the operator inversion in the definition \eqref{eq:defDtNinftrans} of $\trinf{\DtN_{\infty}}$, we introduce $\lambda \in  H^{-\frac{1}{2}}({\trinf{\Gamma_\infty}})$ such that
\begin{equation}
\lambda:= \left(\trinf{S}\right)^{-1}\circ\left(\left(\trinf{D}-\frac{1}{2}I\right)f\right),
\end{equation}
so that
\begin{equation}
\label{eq:DtNinf2courants}
\left\{
\begin{aligned}
\dsp \trinf{\DtN_{\infty}}\left(f\right)&= - \trinf{N}f+\left(\frac{1}{2}I-\trinf{D^*}\right)\lambda,\\
\dsp \left(\trinf{D}-\frac{1}{2}I\right)f-\trinf{S}\lambda&= 0.
\end{aligned}
\right.
\end{equation}

Plugging \eqref{eq:DtNinf2courants} into \eqref {eq:I_infty} and using the expression \eqref{eq:surf_int_mod_bonrepere_v1} in \eqref{eq:varform_intdomain_notransf} leads to the following weak formulation for the coupled problem:
find $\left(f,\lambda\right)\in H^1(\trinf{\Omega^i})\times H^{-\frac{1}{2}}({\trinf{\Gamma_\infty}})$ 
such that $\forall \left(f^t,\lambda^t, \right)\in H^1(\trinf{\Omega^i})\times H^{-\frac{1}{2}}({\trinf{\Gamma_\infty}})$,
\begin{equation}
\label{eq:coupledsys}
\left\{
\begin{aligned}
&\trinf{a}({f},f^t)+J_\infty\int_{\trinf{\Gamma_\infty}}\trinf{N}f\,\bar{{f}^t} 
    + J_\infty\int_{\trinf{\Gamma_\infty}}\left(\trinf{D^*}-\frac{1}{2}I\right)\lambda \, \bar{{f}^t} 
- \frac{\rho_M}{\rho_\infty} \sum_{(\mtimesn)} \varsigma_{\mn} Y_{\mn}^- \bar{\gamma_{\mn}}\\
&\qquad= \frac{\rho_M}{\rho_\infty} \sum_{(\mtimesn)} \alpha_{\mn} \left(Y_{\mn}^+ - Y_{\mn}^-\right) \bar{\gamma_{\mn}},\\
& J_\infty\int_{\trinf{\Gamma_\infty}} \left(\trinf{D}-\frac{1}{2}I\right)f \, \bar{{\lambda}^t}
            -J_\infty\int_{\trinf{\Gamma_\infty}} \trinf{S}\lambda \, \bar{{\lambda}^t}
            = 0.
\end{aligned}
\right.
\end{equation}
with $\varsigma_{\mn}$ and $\gamma_{\mn}$ such that $\dsp \sum_{(\mtimesn)} \left| \varsigma_{\mn}\right|^2  \left| Y^{\pm}_{\mn}\right| < \infty$ and 
 $\dsp \sum_{(\mtimesn)} \left| \gamma_{\mn}\right|^2  \left| Y^{\pm}_{\mn}\right| < \infty$ and with $\trinf{a}(f,f^t)=a(\varphi,\varphi^t)$, where $a$ is defined in \eqref{eq:define_variational_volume_operator}.

\section{Methodologies for the numerical resolutions}
\label{sec:meth_num}
The weak formulation \eqref{eq:coupledsys} has to be solved numerically.
To do so, we first introduce an unstructured volumic mesh $\mathcal{V}_h$ of the domain $\trinf{\Omega^i}$ made of tetrahedrons.
The surfacic meshes $\mathcal{S}_{h,M}$ and $\mathcal{S}_{h,\infty}$
are obtained as the boundary faces of $\mathcal{V}_h$ associated to $\trinf{\Gamma_M}$ and $\trinf{\Gamma_\infty}$
respectively. We denote $\mathcal{V}_h^1$ and $\mathcal{S}_{h,M}^1$ the finite element
spaces $\mathbb{P}^1$ on respectively $\mathcal{V}_h$ and $\mathcal{S}_{h,M}$, and
$\mathcal{S}_{h,\infty}^0$ the finite element space $\mathbb{P}^0$ on $\mathcal{S}_{h,\infty}$.
To introduce a numerical approximation, we have to consider a finite number of modes. 
We consider then $M_{\rm tot}^{\rm inc}$ incident modes and $M_{\rm tot}^{'\rm diff}$ diffracted modes, with $M_{\rm tot}^{'\rm diff}\geq M_{\rm tot}^{\rm inc}$.

We obtain the following discrete conforming approximation of $\eqref{eq:coupledsys}$: 
Find $\left(f_h,\lambda_h\right)\in \mathcal{V}_h^1\times \mathcal{S}_{h,\infty}^0$ 
such that $\forall \left(f_h^t,\lambda_h^t, \right)\in \mathcal{V}_h^1\times \mathcal{S}_{h,\infty}^0$,
\begin{equation}
\label{eq:varform_modes_h}
\left\{
\begin{aligned}
&\trinf{a}({f_h},f_h^t)+J_\infty\int_{\trinf{\Gamma_\infty}}\trinf{N}f_h \, \bar{{f_h}^t} 
    + J_\infty\int_{\trinf{\Gamma_\infty}} \left(\trinf{D^*}-\frac{1}{2}I\right)\lambda_h \, \bar{{f_h}^t} 
-   \frac{\rho_M}{\rho_\infty} \sum_{(\mtimesn)}^{M_{\rm tot}^{'\rm diff}}
        \bar{\varsigma_{\mn}} Y_{\mn}^- \bar{\gamma_{\mn}}\\
&\qquad= \frac{\rho_M}{\rho_\infty} \sum_{(\mtimesn)}^{M_{\rm tot}^{'\rm diff}} 
{\alpha_{\mn}} \left(Y_{\mn}^+ - Y_{\mn}^-\right) \bar{\gamma_{\mn}},\\
& J_\infty\int_{\trinf{\Gamma_\infty}} \left(\trinf{D}-\frac{1}{2}I\right)f_h \bar{{\lambda}_h^t}
            -J_\infty\int_{\trinf{\Gamma_\infty}} \trinf{S}\lambda_h \,\bar{{\lambda}_h^t}
            = 0,
\end{aligned}
\right.
\end{equation}
with $\varsigma_{\mn}$ and $\gamma_{\mn}$ such that $\dsp \sum_{(\mtimesn)} \left| \varsigma_{\mn}\right|^2  \left| Y^{\pm}_{\mn}\right| < \infty$ and 
 $\dsp \sum_{(\mtimesn)} \left| \gamma_{\mn}\right|^2  \left| Y^{\pm}_{\mn}\right| < \infty$. Notice that since $M_{\rm tot}^{'\rm diff}\geq M_{\rm tot}^{\rm inc}$, then some $\alpha_{\mn}$ are zero.

Let $(\theta_i)_{1\leq i\leq p}$ and $(\psi_i)_{1\leq i\leq q}$ denote finite element bases for $\mathcal{V}_h^1$ and $\mathcal{S}_{h,\infty}^0$ respectively.
The decomposition of $f_h\in\mathcal{V}_h^1$ and $\lambda_h\in \mathcal{S}_{h,\infty}^0$ on these bases
are written in the form $f_h=\sum_{i=1}^p {f}_i \theta_i$ and $\lambda_h=\sum_{i=1}^q {\lambda}_i \psi_i$.
Let
\begin{equation}
u:=\begin{pmatrix}
\begin{array}{c}
({f}_i)_{~{1\leq i\leq p}} \\
({\lambda}_i)_{~{1\leq i\leq q}}
\end{array}
\end{pmatrix},\quad
b:=\begin{pmatrix}
\begin{array}{c}
\dsp \left(\frac{\rho_M}{\rho_\infty} \sum_{(\mtimesn)}^{M_{\rm tot}^{'\rm diff}} 
{\alpha_{\mn}} \left(Y_{\mn}^+ - Y_{\mn}^- \right) \int_{\trinf{\Gamma_M}} {\trinf{v_{\mn}}} \, \bar{\theta_i} \right)_{~{1\leq i\leq p}}\\
\left(0\right)_{~{1\leq i\leq q}}
\end{array}
\end{pmatrix},
\end{equation}
and
\begin{equation}
\label{eq:big_matrix}
A:=\begin{pmatrix}
\begin{array}{c|c}
C_{ij}&
\dsp J_\infty\int_{\trinf{\Gamma_\infty}} \left(\trinf{D^*}-\frac{1}{2}I\right)\psi_j \, \bar{\theta_i} \\
\hline
\dsp J_\infty\int_{\trinf{\Gamma_\infty}} \left(\trinf{D}-\frac{1}{2}I\right)\theta_j \, \bar{\psi_i}&
\dsp -J_\infty\int_{\trinf{\Gamma_\infty}} \trinf{S}\psi_j \, \bar{\psi_i}
\end{array}
\end{pmatrix},
\end{equation}
with
\begin{equation}
C_{ij}:=
\trinf{a}(\theta_j,\theta_i)+J_\infty\int_{\trinf{\Gamma_\infty}} \trinf{N}\theta_j \, \bar{\theta_i} 
-   \frac{\rho_M}{\rho_\infty} \sum_{(\mtimesn)}^{M_{\rm tot}^{'\rm diff}} 
       Y_{\mn}^-
\int_{\trinf{\Gamma_M}} \trinf{v_{\mn}}\, \bar{\theta_i} \int_{\trinf{\Gamma_M}} \theta_j \, \bar{\trinf{v_{\mn}}}.
\end{equation}
The linear system resulting from \eqref{eq:coupledsys} is
\begin{equation}
Au=b.
\label{eq:matrixform}
\end{equation}

The matrix $A$ contains both dense and sparse blocks. By reordering the unknowns in order to separate the unknowns related to $\trinf{\Omega^i}$ and to $\trinf{\Gamma_\infty}$ (indexed respectively by $V$ and $\Gamma$ in the following), the linear system can be written 
\begin{equation}
 \begin{pmatrix}
  A_{VV} & A_{V\Gamma} \\
  A_{\Gamma V}  & A_{\Gamma\Gamma}\\
 \end{pmatrix}
 \begin{pmatrix}
  u_{V} \\
  u_{\Gamma}\\
 \end{pmatrix}
= 
 \begin{pmatrix}
  b_{V} \\
  b_{\Gamma}\\
 \end{pmatrix}
\label{eq:matrixform_reordered}
\end{equation}
with $A_{VV}$, $A_{V\Gamma}$ and $A_{\Gamma V}$ being sparse matrices and $ A_{\Gamma\Gamma}$ a dense matrix.

To solve \eqref{eq:matrixform_reordered}, a block Gaussian elimination, known as the Schur complement \cite{schur}, is first carried out on the sparse matrices to
eliminate the unknowns of the volume domain. The remaining system is then
\begin{align}%
&\left( A_{\Gamma\Gamma} -  A_{\Gamma V} A_{VV}^{-1}  A_{V\Gamma}  \right) u_\Gamma = b_\Gamma - A_{\Gamma V} A_{VV}^{-1}  b_V \label{eq:linear_system_after_schur}
\end{align}

This linear system can then be solved either with a direct classical $LU$
solver or an iterative solver. Moreover in the case of the iterative solver, the fast multipole method (FMM)
\cite{fmm,  sylvand2002phd, 2006:ParallelSolverLargeProblems} can be used to take into account the dense matrix $A_{\Gamma\Gamma}$
relative to the BEM formulation.  The conditioning of the remaining system is driven by the BEM matrix.  
The classical SPAI preconditioner for BEM formulation \cite{2005:FMM_SPAI_forEM} of matrix $A_{\Gamma\Gamma}$ is then used to improve the converge of the iterative solver for equation \eqref{eq:linear_system_after_schur}.

These resolution strategies are implemented in the \actipole software and make use of the MUMPS solver \cite{MUMPS:1,MUMPS:2} for the 
sparse matrix elimination and of an in-house solver with out-core and MPI functionalities for the remaining system.

\section{Numerical results}
\label{sec:numericalresults}

Even if the test cases are axisymmetric, all the following computations are full-3D computations.
They have been run on a machine with 2x6 intel Xeon ``Westmere'' processors running at 3.06GHz with 72GB RAM per node and infiniband QDR.

\subsection{Zero flow}
The first test case is designed to check the validity of the modeling for a non-uniform medium without flow and of the FEM-BEM coupling.
It consists of a sphere of radius $1$ m centered at the origin with different fluid properties: $\rho_0=\rho_\infty$, $c_0 = 2 c_\infty$ inside the sphere and 
$\rho_\infty=1.2~\textrm{kg.m}^{-3}, c_\infty=340~\textrm{m.s}^{-1}$ outside the sphere. The acoustic potential source is a monopole located outside the sphere at $(0., 0., 2.5)$.
The observable is located outside the sphere at $(0, 1.7, 0)$ and the frequency range of interest is 11 to 500~Hz.

\begin{figure}[htbp]
 \centering
 \includegraphics[width=0.5\textwidth]{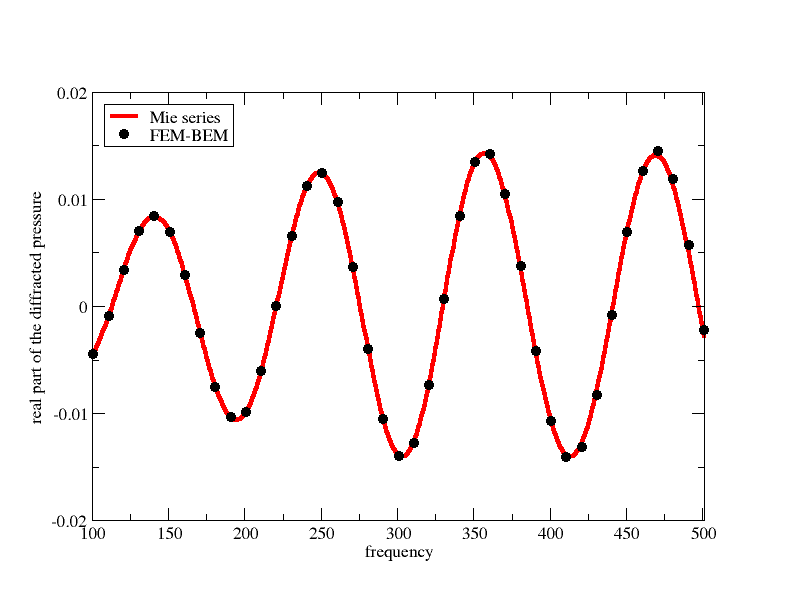}
 % compar_analytic_FEMBEM_BEM_pressure_c2.png: 792x612 pixel, 72dpi, 27.94x21.59 cm, bb=0 0 792 612
 % /home/balin/MULTIPOLE/CodePotentiel/study_boule_IMACS_c
 \caption{Zero flow: Real part of the scattered pressure at the sensor in function of the frequency (validation of the FEM-BEM on the sphere test case).}
 \label{fig:sphere_noflow_valid_fembem}
\end{figure}

The reference result is obtained by a Mie series and the comparison of the scattered pressure with the FEM-BEM solution is visible on Figure \ref{fig:sphere_noflow_valid_fembem}. 
The volumic domain used for the FEM-BEM computation is a sphere of radius 1.07 m to ensure the continuity of the speed of sound
at $\Gamma_\infty$. The mesh has an average edge length of 85 mm ($\lambda/8$ for 500 Hz, the highest frequency) and contains
19,494 dofs (80,616 tetrahedrons and 4,672 triangles on $\Gamma_\infty$). 
The relative error on the module of the total pressure is between $7\times 10^{-4}$ (at 100 Hz) and $3 \times 10^{-2}$
(at 500 Hz).

\subsection{Modal test case without flow}
\label{sec:modalwnoflow}

\begin{figure}[htbp]
 \centering
 \includegraphics[width=0.5\textwidth]{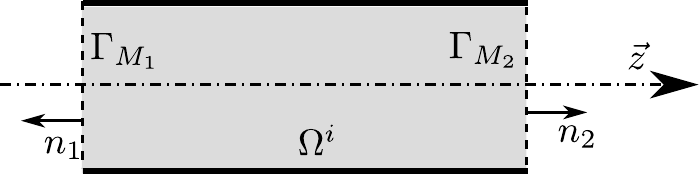}
 % dessin_biface.pdf: 201x50 pixel, 72dpi, 7.09x1.76 cm, bb=0 0 201 50
 \caption{Geometry of the modal test case.}
 \label{fig:dessin_biface}
\end{figure}

Consider a modal cylindrical duct of length $L=1$ m and radius $R=0.25$ m without flow (Figure \ref{fig:dessin_biface}). % (Fig.~\ref{fig:fem_bem_config_modal_duct}).
The frequency of the source is 2,040 Hz. %This problem has been solved with the classical formulation with only BEM in pressure of \actipole and with the new coupled potential formulation presented here.
Three  kinds of computation have been performed with \actipole:
 classical  BEM formulation,
 full FEM formulation,
 coupled FEM-BEM formulation in the same domain of computation. % between the FEM and the BEM domains.
The iterative solver has been chosen with a tolerance on the residual of $10^{-8}$. 
Comparisons have been carried out on the transmission coefficients between $\Gamma_{M_1}$ and $\Gamma_{M_2}$ of the mode $m=0,\, n=1$
(Table~\ref{tab:error_coeff_transm_mode01_noflow})
and their theoretical values.

\begin{table}[htbp]
\centering
\begin{tabular}{|l|c|c|c|c|}
\hline
 \multirow{2}{*}{Mean edge }       &      20~mm      &      16~mm        &        11m~m     &       8~mm  \\
                                   & ($\lambda/8$) &( $\lambda/10$)  & ($\lambda/15$) & ($\lambda/20$) \\\hline
BEM only, FMM     & $9.6 \times 10^{-3}$ & $3.1 \times 10^{-3}$ & $1.2 \times 10^{-3}$ & $2.4 \times 10^{-4}$\\\hline %0.003453 & 0.001346 \\
FEM-BEM, FMM   & 0.28               & 0.19                 & 0.08                 & $4.6 \times 10^{-2}$\\\hline
FEM only, iterative  & 0.55               & 0.36                 & 0.16                 & 0.09\\\hline
\end{tabular}
 \caption{Relative error on the transmission coefficient for the mode $(0,1)$.
\label{tab:error_coeff_transm_mode01_noflow}}
\end{table}

As expected, due to finite element dispersion, the mesh must be finer in the FEM part than in the BEM part to have an acceptable error. 
Moreover the error obtained here by the FEM on the transmission coefficient has a linear dependency on the length of the FEM domain and a square dependency on the size of the elements.

\subsection{Modal test case with an uniform flow}

Consider the previous modal test case with an uniform flow in the direction of the duct, with a Mach number of 0.6.
The frequency of the source is chosen such that the mean number of elements per wavelength after the \PGtransf remains the
same as the previous configuration meshes (without flow) of Section \ref{sec:modalwnoflow}. We recall that the \PGtransf consists in a space dilatation \eqref{eq:def_PGtransformation} and a frequency change \eqref{eq:pb_sep_exterior_PGtransformed}. 
The frequency is then 1305~Hz.

The results on the transmission coefficient for the modes $(0,1)$ are compared with their theoretical values.
The coefficients for both a propagation with the flow (emission on $\Gamma_{M_1}$) and against the flow (emission on $\Gamma_{M_2}$)
are considered.
A computation has also been added to the previous tested configurations. It consists in the case of the full FEM model without any flow in the exterior external domain ($M_\infty=0$). The results are presented on Table~\ref{tab:error_coeff_transm_mode01_againstflowM06}.

\begin{table}[htbp]
\centering
{\itshape Propagation with the flow (emission on modal surface $\Gamma_{M_1}$, reception on $\Gamma_{M_2}$)} \\
\begin{tabular}{|ll|c|c|c|c|}
\hline
 \multirow{2}{*}{Mean edge }       &&      20~mm      &      16~mm        &        11~mm     &       8~mm  \\
                                   && ($\lambda'/8$) &( $\lambda'/10$)  & ($\lambda'/15$) & ($\lambda'/20$) \\\hline
BEM only, FMM& $M_M=M_\infty=0.6$  
                  & $1.2 \times 10^{-2}$& $4.2 \times 10^{-3}$ & $1.6 \times 10^{-3}$ & $5.5 \times 10^{-4}$\\\hline 
FEM-BEM, FMM& $M_M=M_\infty=0.6$   
                  & 0.31               & 0.21                 & 0.09                 & 0.05 \\\hline
FEM only, iterative& $M_M=M_\infty=0.6$ 
                  & 0.61             & 0.40                 & 0.17                 & 0.10\\\hline
FEM only, iterative& $M_M=0.6$, $M_\infty=0$  
                  & 0.0163             & 0.010                 & 0.0045                 & $2.5 \times 10^{-3}$\\\hline
\end{tabular}

~\\
{\itshape Propagation against the flow (emission on modal surface $\Gamma_{M_2}$, reception on $\Gamma_{M_1}$)} \\
\begin{tabular}{|ll|c|c|c|c|}
\hline
 \multirow{2}{*}{Mean edge }       &&      20~mm      &      16~mm        &        11~mm     &       8~mm  \\
                                   && ($\lambda'/8$) &( $\lambda'/10$)  & ($\lambda'/15$) & ($\lambda'/20$) \\\hline
BEM only, FMM& $M_M=M_\infty=0.6$  
                  & $1.2 \times 10^{-2}$& $4.2 \times 10^{-3}$ & $1.6 \times 10^{-3}$ & $5.6 \times 10^{-4}$\\\hline 
FEM-BEM, FMM& $M_M=M_\infty=0.6$   
                  & 0.31               & 0.21                 & 0.09                 & 0.05 \\\hline
FEM only, iterative& $M_M=M_\infty=0.6$ 
                  & 0.61             & 0.40                 & 0.17                 & 0.10\\\hline
FEM only, iterative& $M_M=0.6$, $M_\infty=0$  
                  & 2.00             & 1.60                 & 0.48                 & 0.02\\\hline
\end{tabular}
 \caption{Relative error on the transmission coefficient for the mode $(0,1)$, $M_M=0.6$
\label{tab:error_coeff_transm_mode01_againstflowM06}}
\end{table}

We can see in the first three lines of each array of Table \ref{tab:error_coeff_transm_mode01_againstflowM06} that the errors are very similar to the previous case without flow. The small differences are due to the fact that even if the size of mesh is adapted to the \PGtransf, the mesh is slightly distorted by the transformation (the dilatation factor in the direction of the flow is 1.25).  
Moreover, with the \PGtransf the error is identical whether the mode propagates with or against the flow.
As expected, if the flow outside the duct is zero (fourth line of the arrays of Table \ref{tab:error_coeff_transm_mode01_againstflowM06}), the accuracy is different whether  the wave propagates with or against the flow inside the duct (with respectively large or small equivalent wavelengths). Then if the potential flow is close to the flow at infinity, by using the \PGtransf, a better control on the mesh size and the accuracy can be obtained.

\subsection{Toward engineering applications}

\subsubsection{Rigid sphere into a potential flow}

\begin{figure}[htbp]
 \centering
 \includegraphics[width=0.4\textwidth]{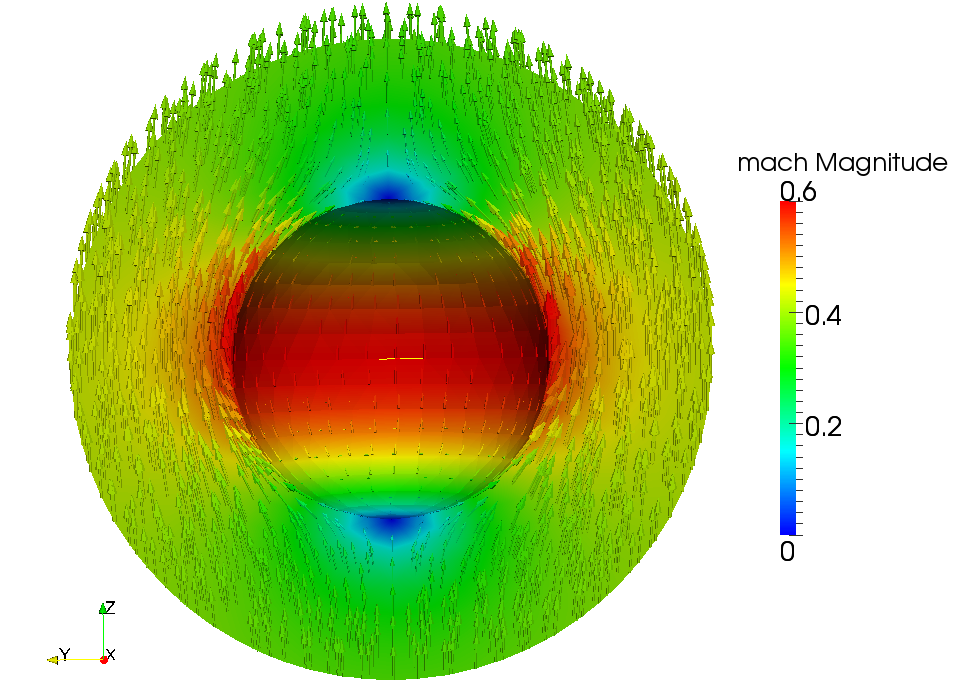}
 % visu_flowMachSphere_cut3D.png: 956x826 pixel, 72dpi, 33.73x29.14 cm, bb=0 0 956 826
 \caption{Analytic flow computed around the sphere.}
 \label{fig:visu_flowMach3d_sphere}
\end{figure}

The next test case is the case of a rigid sphere of radius $R_s=0.6$ m in a flow.
The acoustic source consists of a potential monopole at a frequency of $1133$ Hz
and a distance of $1.2$ m from the surface of the sphere, defined by
\begin{equation}
\varphi^{\inc} =  \frac{e^{i k r}}{4 \pi r}.
 \label{eq:def_potential_monopole}
\end{equation}
We consider two configurations:
\begin{enumerate}
 \item an uniform flow defined by $\vec{M_\infty} = 0.4 \vec{e_z}$. There is no interior domain $\Omega^i$, and the boundary
conditions at $\Gamma$ are clearly violated,
 \item an incompressible analytic potential flow around the sphere (Equation \eqref{eq:potentialflow_sphere} for $r<R_\infty$) combined with an uniform flow far from the sphere ($\vec{M_\infty} = 0.4 \vec{e_z}$ for $r\ge R_\infty$) with a supposed continuity of the flow at the interface. For that we choose $ R_\infty = 2 R_s = 1.2$ m (Figure \ref{fig:visu_flowMach3d_sphere}).
\end{enumerate}

In spherical coordinates, the potential flow in $\Omega^i$ in spherical coordinates is such that
\begin{equation}
\vec{M_{0}} (r,\theta, \phi) = M_\infty \cos(\theta) \left[ 1 - \left( \frac{R_s}{r} \right)^3 \right]  \vec{e_r} 
                             - M_\infty \sin(\theta)  \left[ 1 + \frac{1}{2} \left( \frac{R_s}{r} \right)^3  \right] \vec{e_\theta}.
\label{eq:potentialflow_sphere}
\end{equation}
The flow is then tangent to $\Gamma$, but the continuity condition of the flow through $\Gamma_\infty$ is not strictly obtained for a finite value of $R_\infty$.

A mesh with an average edge length of 25 mm is used. That represents $6.7\times10^{5}$ degrees of freedom in the volume and $10^{5}$ on the surface. The computation took 7 h on 60 processors for the direct solver and  2 h on 24 processors and 568 iterations for the FMM solver for an achieved residual of $10^{-6}$.

\begin{figure}[htbp]
 \centering
 \includegraphics[width=0.2305\textwidth]{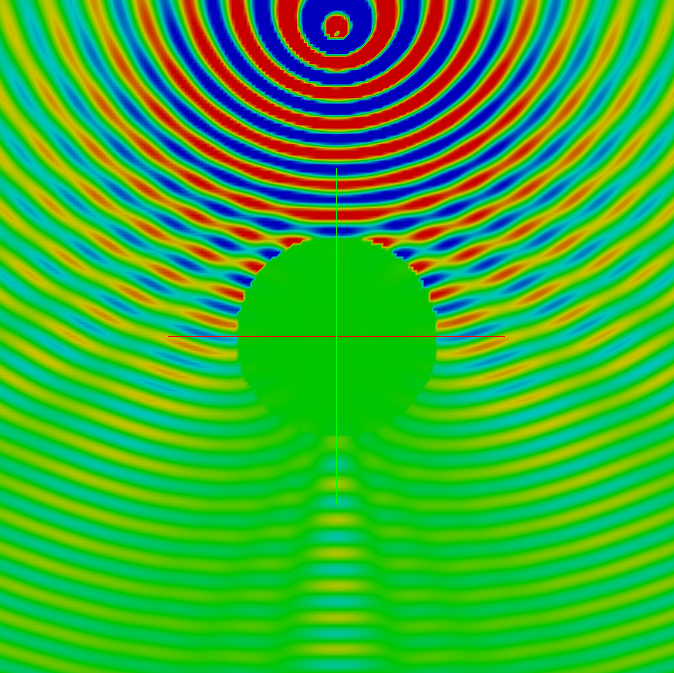}
 % sphere_pressure.png: 482x607 pixel, 72dpi, 17.00x21.41 cm, bb=0 0 482 607
 \includegraphics[width=0.23\textwidth]{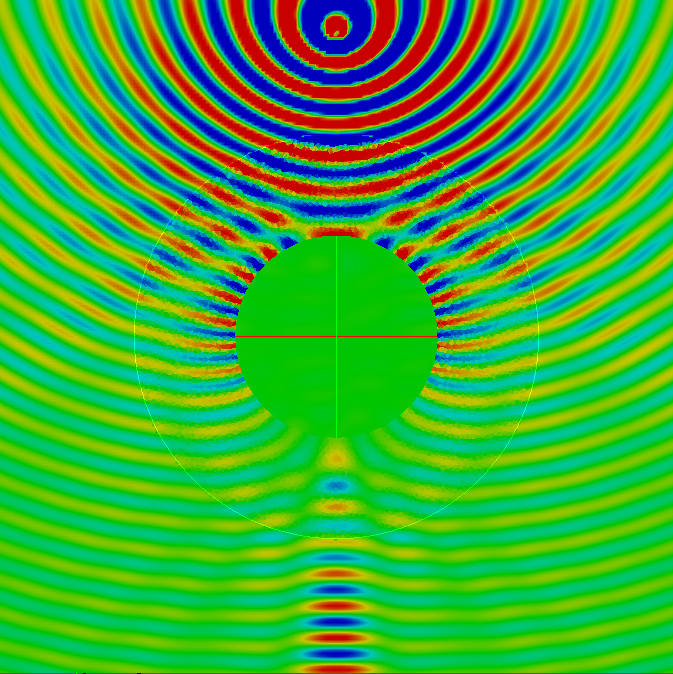}
 \caption[Acoustic total pressure around a sphere with a potential flow]{Rigid sphere: real part of the total acoustic pressure for an emitter located at $(0.,0.,1.8)$
 (\textit{left}: uniform flow BEM, \textit{right}: potential flow FEM-BEM).}
 \label{fig:potentialflow_sphere}
\end{figure}

\begin{figure}[htbp]
 \centering

 \includegraphics[width=0.36\textwidth]{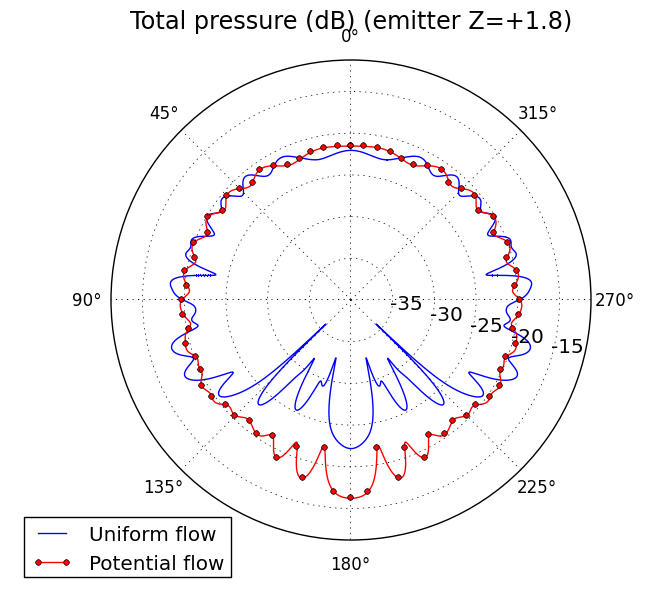} 
 \includegraphics[width=0.36\textwidth]{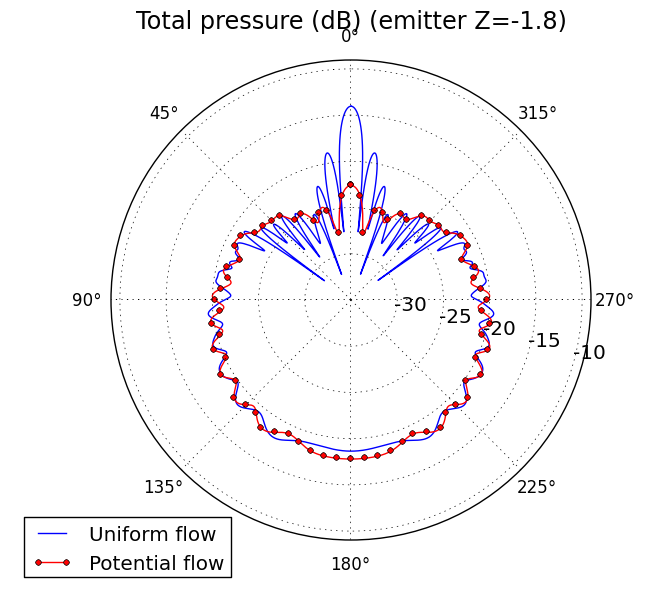} 

 \includegraphics[width=0.36\textwidth]{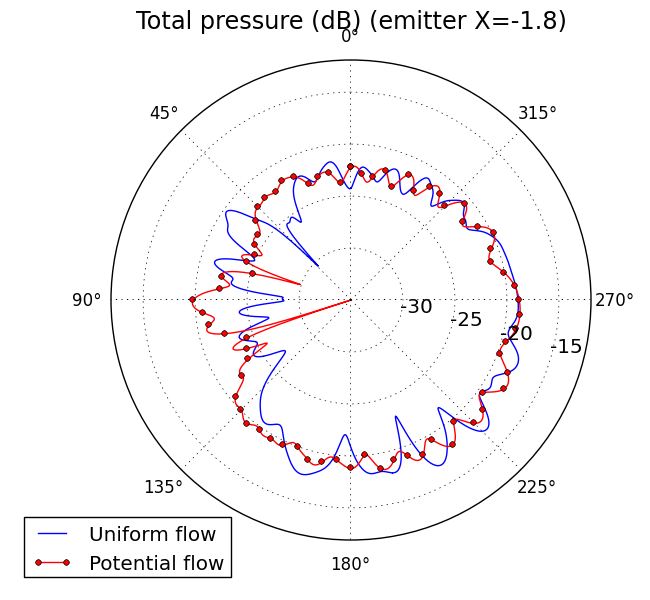} '
\caption{Rigid sphere: acoustic total pressure in dB radiated on a circle at $r=10$~m for different positions of the emitter
\label{fig:Pradiated_sphere_10m}.}
\end{figure}

Figure~\ref{fig:potentialflow_sphere} illustrates that the presence of the potential flow around the sphere has modified the acoustic potential map. Local acoustic velocity and pressure magnitude have increased, as well as its magnitude in the shadow zone. This is also visible on Figure \ref{fig:Pradiated_sphere_10m}, that shows some radiation patterns for the total pressure on a circle of radius 10 m for 3 positions of the emitter ( $(0.,0.,1.8)$, $(0.,0.,-1.8)$ and $(-1.8,0.,0)$ respectively). 
Figure~\ref{fig:potentialflow_sphere} shows that first the radiation pattern is modified by the hypothesis on the flow, and second 
that this modification is different whether the acoustic waves propagate with or against the flow, with respectively lower and higher level of pressure for the potential model in the shadow region.

\subsubsection{Simplified engine}

\begin{figure}[htbp]
 \centering
 \includegraphics[width=0.45\textwidth]{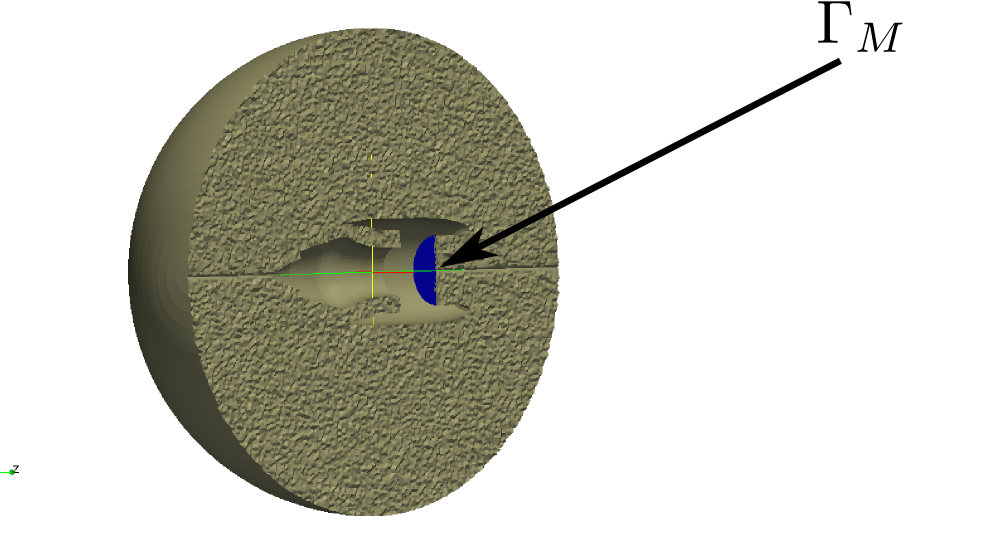}
 % visu_mesh_volumic.png: 0x0 pixel, 300dpi, 0.00x0.00 cm, bb=
 \includegraphics[width=0.45\textwidth]{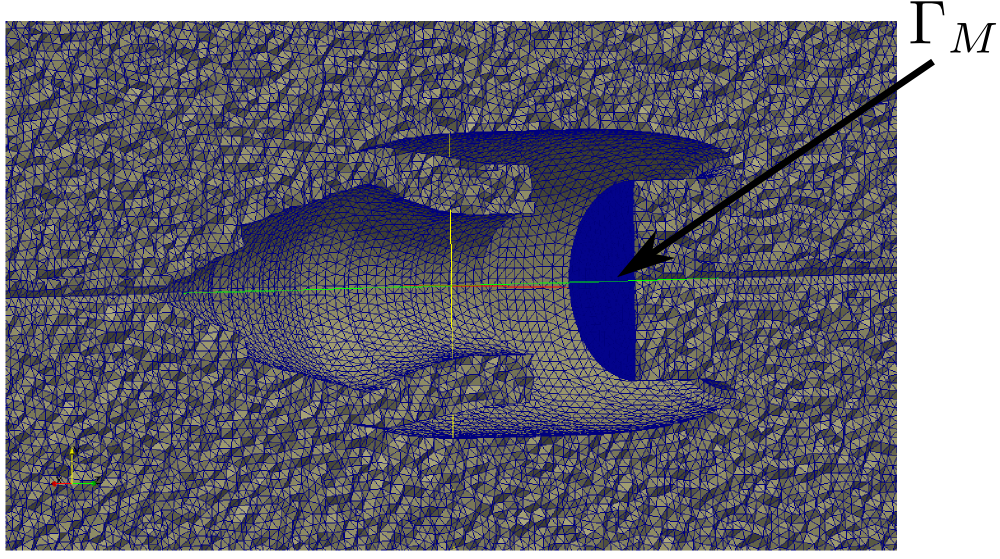}
 % visu_mesh_volumic.png: 0x0 pixel, 300dpi, 0.00x0.00 cm, bb=
 \caption{Cut of the mesh of the simplified engine.}
 \label{fig:model_engine}
\end{figure}

The next test case is more realistic. It consists of a simplified engine with modal
surfaces orthogonal to $\vec{e_z}$ to model the upstream and downstream fans (see Figure \ref{fig:model_engine}).
The far field flow is defined by $\vec{M_\infty} = - 0.3 \vec{e_z}$. Three different configurations
are considered: an uniform flow defined by $\vec{M_\infty}$ and potential flows computed such that the Mach number at the
upstream modal surface $\Gamma_M$ is 0.3 and 0.42.

\begin{figure}[htbp]
\center
 \includegraphics[width=0.45\textwidth]{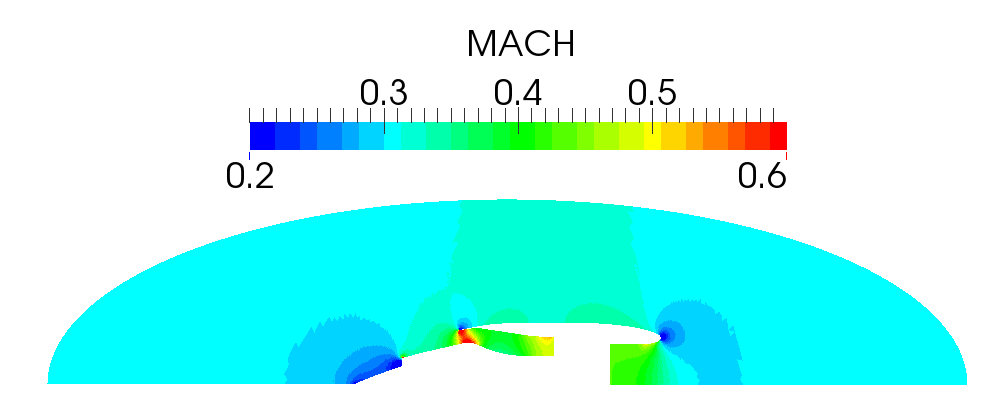}
 \includegraphics[width=0.45\textwidth]{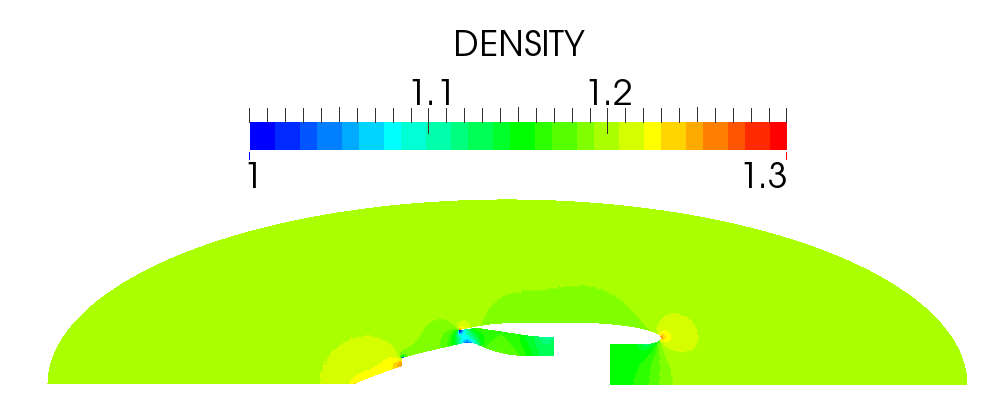}
 % EcoultMoteur_M0.2Moteur1_M0.1infini_mach.png: 1478x826 pixel, 72dpi, 52.14x29.14 cm, bb=0 0 1478 826
 \caption{Simplified engine: computed potential flow such that $M_M=0.42$ and $M_\infty=0.3$
(\textit{left}: Mach number, \textit{right}: density).}
 \label{fig:engine_potential_flow}
\end{figure}

First, the potential flow is computed using an in-house software based on a fixed-point algorithm \cite{Pe75, duprey}.
The potential flow obtained when $M_M = 0.42$ at the upstream modal surface is plotted on Figure \ref{fig:engine_potential_flow}.

We now consider the upstream fan modal source model at the frequency of 200 Hz. The mean size of the mesh elements is 83 mm.
The model contains $1.2~\times 10^6$ dofs and $11.8~\times 10^6$ tetrahedrons. At this frequency and for this flow, there are three to four propagating modes.
To be compared, the intensity on each mode is set to 100 dB, following Morfey's convention \cite{Morfey1971159}.

\begin{figure}[htbp]
 \centering
\begin{tabular}{ccc}%c}
\rotatebox{90}{\small{~Uniform flow}} &
 \includegraphics[width=0.3\textwidth]{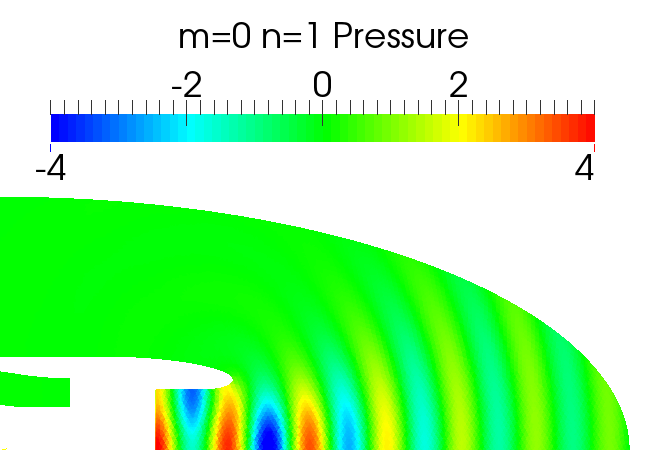} & 
 \includegraphics[width=0.3\textwidth]{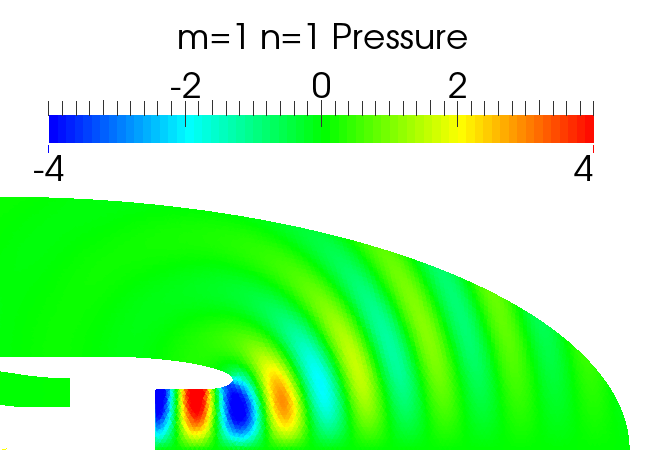} \\
\rotatebox{90}{\small{~Potential flow}} &
 \includegraphics[width=0.3\textwidth]{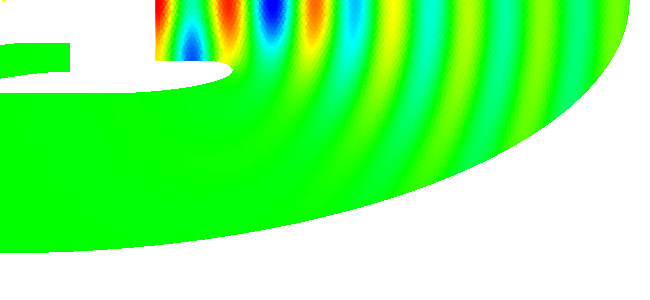} & 
 \includegraphics[width=0.3\textwidth]{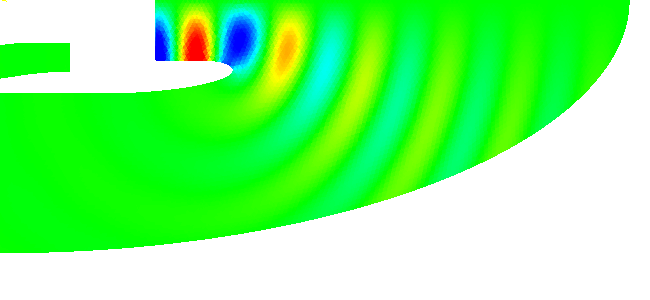} \\ 
\end{tabular}

 % nicevisu_Pre_mode_0_1_uniforme_Mm02.png: 1027x826 pixel, 72dpi, 36.23x29.14 cm, bb=0 0 1027 826
\caption{Simplified engine: comparison of the pressure for the uniform model and the potential model with $M_\infty=M_M=0.3$. %for the propagating modes
\label{fig:engine_Pre_modes_compar_uniform_potential_1}}
\end{figure}

\begin{figure}[htbp]
 \centering
\begin{tabular}{ccc}%c}
\rotatebox{90}{\small{~Uniform flow}} &
 \includegraphics[width=0.3\textwidth]{nicevisu_Pre_mode_0_1_uniforme_Minf03_Mm03.png} & 
 \includegraphics[width=0.3\textwidth]{nicevisu_Pre_mode_1_1_uniforme_Minf03_Mm03.png} \\
\rotatebox{90}{\small{~Potential flow}} &
 \includegraphics[width=0.3\textwidth]{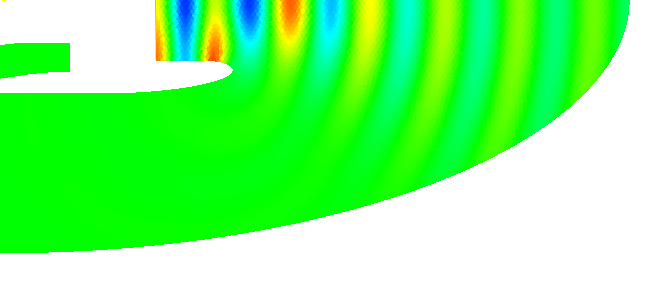} & 
 \includegraphics[width=0.3\textwidth]{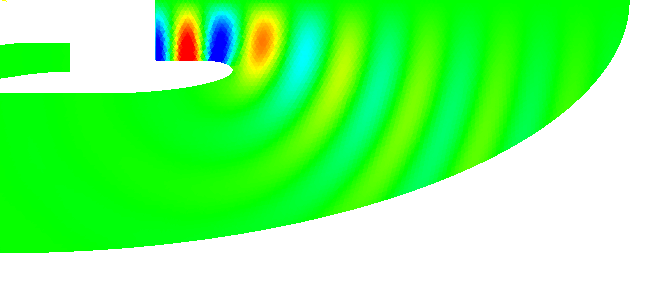} \\ 
\end{tabular}

 % nicevisu_Pre_mode_0_1_uniforme_Mm02.png: 1027x826 pixel, 72dpi, 36.23x29.14 cm, bb=0 0 1027 826
\caption{Simplified engine: comparison of the pressure for the uniform model ($M_\infty=M_M=0.3$) and the potential model ($M_\infty=0.3$ and $M_M=0.42$). %for the propagating modes
\label{fig:engine_Pre_modes_compar_uniform_potential_2}}
\end{figure}

The pressure obtained in the vicinity of the modal surface is shown on Figures \ref{fig:engine_Pre_modes_compar_uniform_potential_1} and \ref{fig:engine_Pre_modes_compar_uniform_potential_2}. For each mode, the top part of the Figure is the pressure obtained with the uniform flow model and the bottom part of the Figure the pressure obtained with the potential model with a Mach number at the modal surface of 0.3 or 0.42. Small variations are observed with the $M_M=0.3$ condition. The differences are higher when the flow at the modal surface is higher.

\begin{figure}[htbp]
 \centering
\begin{tabular}{cc}%c}
 \includegraphics[width=0.36\textwidth]{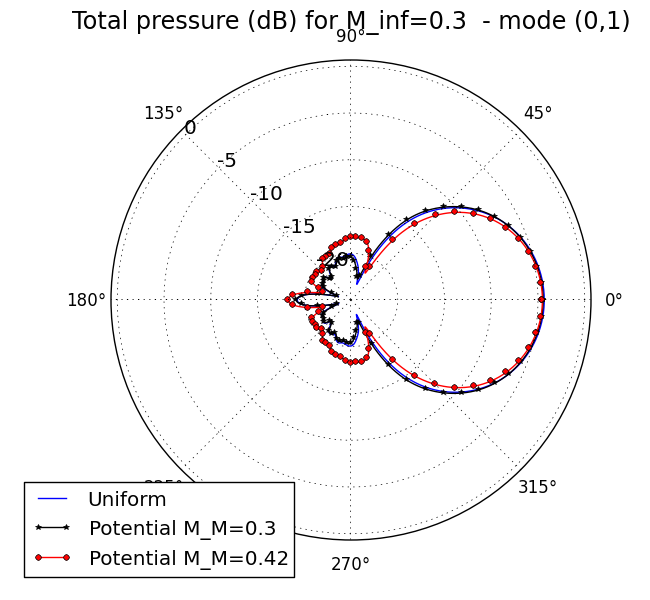}&
 \includegraphics[width=0.36\textwidth]{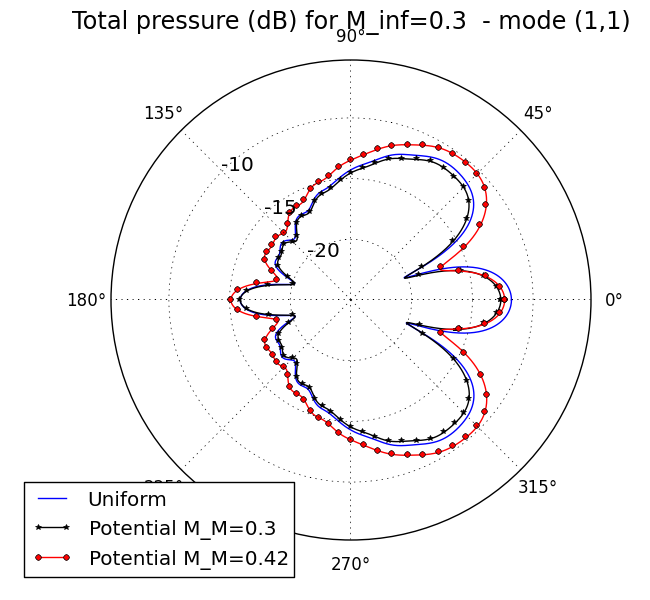}\\%&
\end{tabular}
 % visu_Pre_mode_0_1_uniforme_Mm02.png: 1170x826 pixel, 72dpi, 41.28x29.14 cm, bb=0 0 1170 826
\caption{Simplified engine: pressure in dB on a circle at $r=20$ m for $M_\infty=0.3$  and some values of $M_M$  (mesh size 75mm).
\label{fig:engine_farPdB_modes_compar_uniform_potential}}
\end{figure}
 
Figure \ref{fig:engine_farPdB_modes_compar_uniform_potential} shows the pressure in dB obtained on a circle at a distance of 20~m
from the center of the modal surface, for different values of $M_M$. 
Significant change in the amplitude are obtained for the different modes by taking into account the potential flow. 
For instance for the mode $(1,1)$ and for the same flow at the modal surface, the amplitude predicted for the potential flow is approximately 1~dB lower in the axis direction than the amplitude predicted by the uniform flow model. By increasing the flow through the upstream modal surface, the difference with the uniform flow model is higher and observed for all the radiation directions.

\begin{figure}[htbp]
 \centering
\begin{tabular}{cc}%c}
 \includegraphics[width=0.36\textwidth]{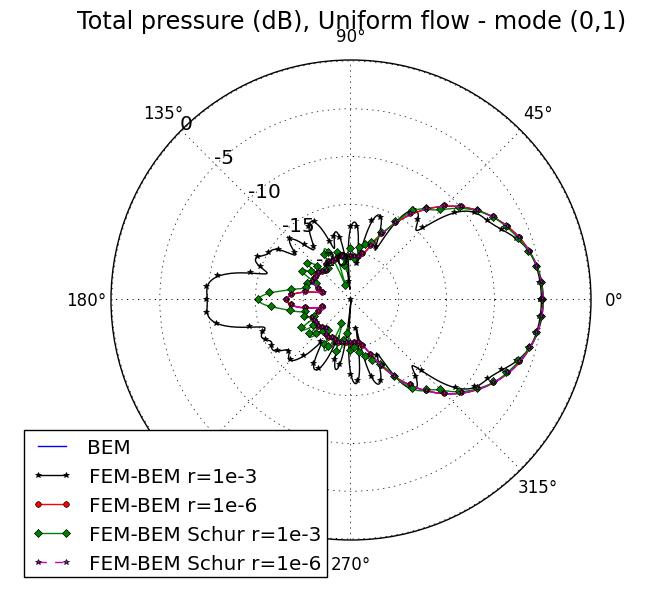}&
 \includegraphics[width=0.36\textwidth]{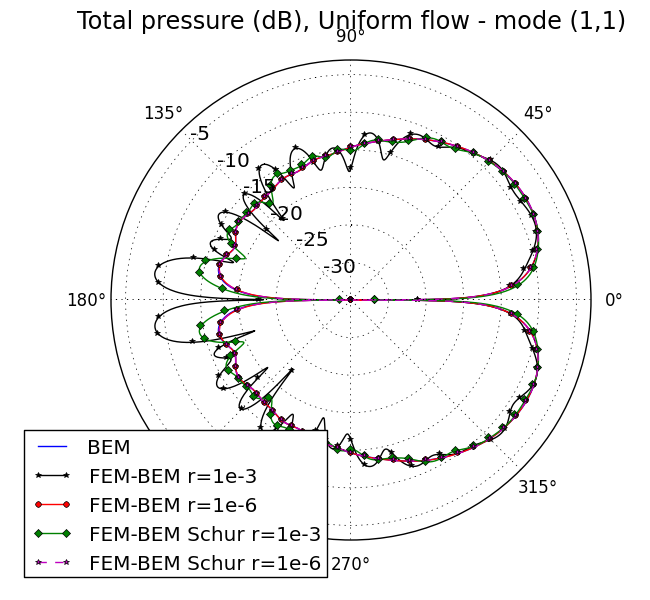}\\%&
\end{tabular}
 % visu_Pre_mode_0_1_uniforme_Mm02.png: 1170x826 pixel, 72dpi, 41.28x29.14 cm, bb=0 0 1170 826
\caption{Simplified engine: influence of the convergence criteria and on the pressure in dB on a circle at $r=20$ m for an uniform flow defined by  $M_M=M_\infty=0.1$.
\label{fig:engine_farPdB_modes_compar_uniform_cv}}
\end{figure}

Figure \ref{fig:engine_farPdB_modes_compar_uniform_cv} illustrates
the influence of the relative residual of the iterative solver on the diffracted pressure field.
Results for relative residuals of $10^{-3}$ and $10^{-6}$, and with or without a Schur complement on the volume part of the matrix are presented. From these results, it appears that a convergence with a tolerance of $10^{-3}$ is not sufficient for a solution without a Schur complement on the volume part of the matrix.

In that case, for a mesh containing $4.7~\times 10^6$ dofs and $25.8~\times 10^6$ tetrahedrons, 
 the computation took 1.5 h on 160 processors and 231 iterations for the FMM solver  without using the Schur complement and 6.5 h on 120 processors and 204 iterations with for an achieved residual of $10^{-6}$.

\section{Conclusion}

In this work, we derived a direct coupling method to compute the acoustic propagation of the noise generated by a turboreactor
in a flow that is potential in a bounded domain containing the object, and uniform elsewhere.
This approach, that decouples the movement of the fluid and the acoustic effect, and uses a simplified model for the flow,
enabled noticeable improvements. 

Mathematical justification of the well posedness of the continuous and discretized formulations can be found in \cite{casenave}.
In particular, the formulation that we used is not invertible at some frequencies of the source. These resonant frequencies
are not physical, since the considered boundary value problem always has a unique solution.
A combined field integral equations method can be derived to recover well-posedness at all the frequencies. Such a method has
been implemented in our code \cite{casenave}.

The method has been implemented for generic 3D configurations.
Complementary tests have to be conducted to catch the limitation of the potential assumption.
However, now that the coupling has been carried out, and considering that the uniform flow assumption is reasonable far from
the object, we can easily enrich the physics of the problem by considering more complex flows in the interior domain
(e.g. rotational flow leading to the Galbrun equation \cite{BM12}), or other boundary condition. %These extensions are well known and only require to work on the finite element part of the resolution.
Measurement campaigns are under analysis, and comparisons with our numerical model will be presented as soon as possible in collaboration with the team responsible for these measurements.

\appendix
 \section{Appendix}

\subsection{Alternative computation of the coupling integral term $I_M$ \eqref{eq:form_var_intmodal}}
\label{sec:appendix_form_modes}

In the same fashion as the exterior problem,
it is possible to transform the modal problem \eqref{eq:modalpb} to recover the classical Helmholtz equation
by introducing a \PGtransf associated to $\vec{M_M}$ (this is the case in \actipole). 

\begin{remark}
In what follows, we note by ~\textbf{$\trM{\cdot}$} the objects and operators transformed by the \PGtransf associated to $\vec{M_M}$ (normals, geometry, derivatives).
\end{remark}

The waveguide is still supposed oriented along a local axis $\vec{e_{z_M}}$ and with the uniform flow $\vec{M_M}$. As $\Gamma_M$ is orthogonal to $\vec{e_{z_M}}$, we suppose $\trM{z_M \phantom{}_{|\Gamma_M}} = 0$. The flow $\vec{M_M}$ is then parallel to $\vec{e_{z_M}}$. 

In $\trM{\Omega_M}$, the transformed acoustic potential is decomposed into an incident and a diffracted potential:
$\trM{\varphi}:=\trM{\varphi^{\rm inc}}+\trM{\varphi^{\rm diff}}$, both satisfying
\begin{equation}
\begin{aligned}
\trM{\Delta} \trM{\varphi^{\rm inc,diff}} + \trM{k_{M}}^2 \trM{\varphi^{\rm inc,diff}} = 0, &&&\textnormal{~in }\trM{\Omega_M}.
\end{aligned}
\label{eq:pb_sep_modal_PGtransformed}
\end{equation}

The incident potential is known% and supposed to satisfy an ingoing Sommerfeld radiation condition at infinity
, whereas the
diffracted potential is unknown. % and supposed to satisfy an outgoing Sommerfeld radiation condition at infinity.
The following decomposition holds \cite{MI68, Br06}:
\begin{equation}
\begin{aligned}
 \trM{\varphi^{\inc}}(\trM{x},\trM{y},\trM{z}) &=\sum_{(\mtimesn) } 
\trM{\alpha_{\mn}}  \, \trM{v_{\mn}}(\trM{x},\trM{y}) \exp\left({i\trM{k_{\mn}^{+}} \trM{z}}\right)\quad\text{in } \trM{\Omega_M},\\
 \trM{\varphi^{\diffr}}(\trM{x},\trM{y},\trM{z})& = \sum_{(\mtimesn)} 
\trM{\beta_{\mn}} \,  \trM{v_{\mn}}(\trM{x},\trM{y})  \exp\left({i\trM{k_{\mn}^{-}} \trM{z}}\right)\quad\text{in } \trM{\Omega_M},
\label{eq:def_decomp_modal_trM}
\end{aligned}
\end{equation}
with $\mn$ a discrete index and where $\trM{\alpha_{\mn}}$ and $\trM{\beta_{\mn}}$ are the incident and diffracted modal coefficients.
The basis functions $\trM{v_{\mn}}$ constitute an orthonormal modal basis function. 

For instance for a cylindrical duct of radius $R$, $\mn$ corresponds to a couple of indices $(m,n) \in (\Z \times \N^*)$ and the functions  $\trM{v_{\mn}}$ are defined in polar coordinates by \cite{MI68, Br06} by
\begin{equation}
\trM{v_{\mn}} \left(\trM{r},\trM{\theta}\right)  =  \trM{v_{m,n}}\left(\trM{r},\trM{\theta}\right) =  \trM{V_{m,n}}\, J_m\left(\frac{r_{m,n}}{R} \, \trM{r}\right) \exp\left({i \, m \, \trM{\theta}}\right)
,
\end{equation}
with $r_{m,n}$ the $n$-th zero of the derivative of $m$-th Bessel function of the first kind $J_m$, $\trM{V_{m,n}}$ the normalization factor such that $\int_{\trM{\Gamma_M}} \trM{v_{\mn}} = 1$, and 
\begin{align}
\trM{ k_{\mn}^{\pm}} = \trM{ k_{mn}^{\pm}} &=  \pm \sqrt{\left(\trM{k_M}\right)^2 - \left(\displaystyle \frac{r_{m,n}}{R}\right)^2}, 
          \quad \text{~ for propagating modes} ~ (\trM{k_{mn}^{\pm}} \in \mathbb{R}),\\ 
\trM{ k_{\mn}^{\pm}} = \trM{ k_{mn}^{\pm}} &=  \pm i \sqrt{\left(\displaystyle \frac{r_{m,n}}{R}\right)^2 -\left(\trM{k_M}\right)^2 }, 
          \quad  \text{~ for evanescent modes} ~ (\trM{k_{mn}^{\pm}} \in \mathbb{C}).
\end{align}
Based on this decomposition, the expression of the Dirichlet-to-Neumann $\trM{\DtN_M}$ operator is \cite{dahi_conditionD2N_modes:2002, PhDlegendre:2003}:
\begin{equation}
 \trM{\vec{\nabla}} \bar{\trM{\varphi}} \cdot \trM{\vec{n}} =
\trM{\DtN_M} \left(  \trM{\varphi} \right)
 := \sum_{(\mtimesn)} 
             \left(\trM{\alpha_{\mn}}  \trM{Y^{+}_{\mn}} +\trM{\beta_{\mn}}  \trM{Y^{-}_{\mn}}\right)v_{\mn}\quad\textnormal{on }\trM{\Gamma_M},
\label{eq:def_operator_DtN_PGM}
\end{equation}
with $\trM{Y^{\pm}_{\mn}}  =  -i\trM{ k^{\pm}_{\mn}}$.
Similarly to Section \ref{sec:transfointeq},
by using the \PGtransf associated to $\vec{M_M}$, we infer 
\begin{align}
\label{eq:surf_int_mod}
 I_M (\varphi^i,\varphi^t)&  =
 - J_M  \frac{\rho_M}{\rho_\infty}  \int_{\trM{\Gamma_M}}  \trM{\vec{\nabla}} \trM{\varphi^{i}} \cdot \trM{\vec{n}}\; \bar{{\trM{\varphi}}^t} \\
& = - J_M  \frac{\rho_M}{\rho_\infty}  \int_{\trM{\Gamma_M}} 
            \trM{\DtN_{M}}\left({\trM{\varphi^i}}\right) \bar{{\trM{\varphi}}^t}
.
\end{align}
Then, using the Dirichlet-to-Neumann operator \eqref{eq:def_operator_DtN_PGM}, we obtain
\begin{equation}
\label{eq:surf_int_mod_v2}
I_M (\varphi^i,\varphi^t)
= - J_M \frac{\rho_M}{\rho_\infty}  \sum_{(\mtimesn)} 
\left[ \trM{\alpha_{\mn}} \trM{Y_{\mn}^+}  
+  \trM{\beta_{\mn}} \trM{Y_{\mn}^-}  \right]
            \int_{\trM{\Gamma_M}} \trM{v_{\mn}} \, \bar{{\trM{\varphi}}^t}.
\end{equation}

However, to carry out a coupling between unknown functions written in the same variables, the integral $I_M$ must be expressed in terms of quantities transformed by the \PGtransf associated to $\vec{M_\infty}$. Using the results obtained in Section~\ref{sec:transfointeq}, we have  that 
\begin{align*}
J_M \int_{\trM{\Gamma_M}}  \trM{g}(\trM{\vec{x}}) \; \bar{\trM{h}}(\trM{\vec{x}}) %d\gamma''
& = 
\int_{\Gamma_M}  \frac{1}{\trM{K_M}(\vec{x})} {g}(\vec{x}) \; \bar{h}(\vec{x}) %d\gamma
= 
\int_{\trinf{\Gamma_M}}  \frac{J_\infty \trinf{K_\infty}(\trinf{\vec{x}})}{\trM{K_M}(\trinf{\vec{x}})}  
        {\trinf{g}}(\trinf{\vec{x}}) \; \overline{\trinf{h}}(\trinf{\vec{x}}), %d\gamma'
\end{align*}
where $\trM{K_M}$ and $\trinf{K_\infty}$ are defined in equation \eqref{eq:Kprime_expression}.

Equation \eqref{eq:surf_int_mod_v2} becomes 
\begin{align}
 I_M (\varphi^i,\varphi^t)
& = - J_\infty \frac{\rho_M}{\rho_\infty}  \sum_{(\mtimesn)} 
\left[ \trM{\alpha_{\mn}} \trM{Y_{\mn}^+}  
+  \trM{\beta_{\mn}} \trM{Y_{\mn}^-}  \right]
            \int_{\trinf{\Gamma_M}} \frac{\trinf{K_\infty}(\trinf{\vec{x}})}{\trM{K_M}(\trinf{\vec{x}})}  
            \trinf{v_{\mn}} \, \bar{\trinf{\varphi}^t}
.
\label{eq:surf_int_mod_v3_TPG}
\end{align}
If $\vec{M_M}$ and $\vec{M_\infty}$ are colinear, there holds
\begin{equation*}
  \frac{\trinf{K_\infty}}{\trM{K_M}}= \sqrt{\frac{{1-M_M^2}}{{1-M_\infty^2}}},
 \end{equation*}
 and otherwise, denoting $\alpha = \vec{M_\infty} \cdot \trinf{\vec{n_M}}$,
\begin{align*}
  \frac{\trinf{K_\infty}}{\trM{K_M}}
    & = \sqrt{ \left[ (1 + C_\infty \alpha^2)(1 + \trM{C_M} M_M^2)\right]^2 + C_\infty^2 \alpha^2 (M_\infty^2 - \alpha^2)}.
\end{align*}

\section*{Acknowledgments}
This work takes part in the AEROSON project, financed by the ANR (French National Research Agency).
The authors would like to express their gratitude to Toufic Abboud (IMACS), 
Patrick Joly (INRIA) and Guillaume Sylvand (Airbus Group) for fruitful discussions
and to Emilie Peynaud and Fran\c cois Madiot for their work during their internship at Airbus Group.

\bibliographystyle{plain}
\bibliography{biblio}

\begin{thebibliography}{10}

\bibitem{AGH99}
S.~Abarbanel, D.~Gottlieb, and J.~S. Hesthaven.
\newblock {Well-posed perfectly matched layers for advective acoustics}.
\newblock {\em Journal of Computational Physics}, 154:266--283, 1999.

\bibitem{AJRT11}
T.~Abboud, P.~Joly, J.~Rodr\`iguez, and I.~Terrasse.
\newblock {Coupling discontinuous Galerkin methods and retarded potentials for
  transient wave propagation on unbounded domains}.
\newblock {\em Journal of Computational Physics}, 230:5877--5907, 2011.

\bibitem{MUMPS:1}
P.~R. Amestoy, I.~S. Duff, J.~Koster, and J.-Y. L'Excellent.
\newblock A fully asynchronous multifrontal solver using distributed dynamic
  scheduling.
\newblock {\em SIAM Journal on Matrix Analysis and Applications}, 23(1):15--41,
  2001.

\bibitem{MUMPS:2}
P.~R. Amestoy, A.~Guermouche, J.-Y. L'Excellent, and S.~Pralet.
\newblock Hybrid scheduling for the parallel solution of linear systems.
\newblock {\em Parallel Computing}, 32(2):136--156, 2006.

\bibitem{Amiet}
R.~Amiet and W.~R. Sears.
\newblock The aerodynamic noise of small-perturbation subsonic flows.
\newblock {\em Journal of Fluid Mechanics}, 44:pp. 227--235, 1970.

\bibitem{Astley1986445}
R.J. Astley and J.G. Bain.
\newblock A three-dimensional boundary element scheme for acoustic radiation in
  low mach number flows.
\newblock {\em Journal of Sound and Vibration}, 109(3):445 -- 465, 1986.

\bibitem{inteqconvHelmholtz}
M.~{Beldi} and F.~S. {Monastir}.
\newblock Resolution of the convected {H}elmholtz's equation by integral
  equations.
\newblock {\em The Journal of the Acoustical Society of America}, 103:2972,
  1998.

\bibitem{Be94}
J.~P. B\'erenger.
\newblock {A perfectly matched layer for the absorption of electromagnetic
  waves}.
\newblock {\em Journal of Computational Physics}, 114:185--200, 1994.

\bibitem{BM91}
J.~Bielak and R.~C. Mac~Camy.
\newblock {Symmetric Finite Element and Boundary Integral Coupling Methods for
  Fluid-Solid Interaction}.
\newblock {\em Quarterly of Applied Mathematics}, 49:107--119, 1991.

\bibitem{dahi_conditionD2N_modes:2002}
A.-S. Bonnet-Ben~Dhia, L.~Dahi, E.~Luneville, and V.~Pagneux.
\newblock Acoustic diffraction by a plate in a uniform flow.
\newblock {\em Mathematical Models and Methods in Applied Sciences},
  12(05):625--647, 2002.

\bibitem{BM12}
A.-S. Bonnet-Ben~Dhia, J.-F. Mercier, F.~Millot, S.~Pernet, and E.~Peynaud.
\newblock {Time-Harmonic Acoustic Scattering in a Complex Flow: a Full Coupling
  Between Acoustics and Hydrodynamics}.
\newblock {\em Communications in Computational Physics}, 11(2):555--572, 2012.

\bibitem{Br11}
E.~J Brambley.
\newblock Well-posed boundary condition for acoustic liners in straight ducts
  with flow.
\newblock {\em AIAA journal}, 49(6):1272--1282, 2011.

\bibitem{Br06}
M.~Bruneau and T.~Scelo.
\newblock {\em Fundamentals of Acoustics}.
\newblock ISTE, 2006.

\bibitem{BM71}
A.~J. Burton and G.~F. Miller.
\newblock The application of integral equation methods to the numerical
  solution of some exterior boundary-value problems.
\newblock {\em Proceedings of the Royal Society of London. A. Mathematical and
  Physical Sciences}, 323(1553):201--210, 1971.

\bibitem{2005:FMM_SPAI_forEM}
B.~Carpentieri, I.~S. Duff, L.~Giraud, and G.~Sylvand.
\newblock Combining fast multipole techniques and an approximate inverse
  preconditioner for large electromagnetism calculations.
\newblock {\em SIAM Journal on Scientific Computing}, 27:774--792, 2005.

\bibitem{casenave}
F.~Casenave, A.~Ern, and G.~Sylvand.
\newblock Coupled {BEM-FEM} for the convected {H}elmholtz equation with
  non-uniform flow in a bounded domain.
\newblock {\em Journal of Computational Physics}, 257, Part A(0):627--644,
  2014.

\bibitem{CHAPMAN}
C.~J. Chapman.
\newblock Similarity variables for sound radiation in a uniform flow.
\newblock {\em Journal of Sound and Vibration}, 233(1):157 -- 164, 2000.

\bibitem{Ci78}
P.~G. Ciarlet.
\newblock {\em { The Finite Element Method for Elliptic Problems}}.
\newblock North Holland, Amsterdam, 1978.

\bibitem{Costabel}
M.~Costabel.
\newblock {\em Symmetric methods for the coupling of finite elements and
  boundary elements}.
\newblock Preprint. Fachbereich Mathematik. Technische Hochschule Darmstadt.
  Fachber., TH, 1987.

\bibitem{Cr73}
A.~Craggs.
\newblock {An acoustic finite element approach for studying boundary
  flexibility and sound transmission between irregular enclosures}.
\newblock {\em Journal of Sound and Vibration}, 30:343--357, 1973.

\bibitem{justif_actipole_2011}
A.~Delnevo.
\newblock {Code ACTI3S harmonique : Justification Math\'ematique}.
\newblock Technical report, EADS CCR, 2001.

\bibitem{delnevo}
A.~Delnevo.
\newblock {Code ACTI3S, Justifications Math\'ematiques : partie II, pr\'esence
  d'un \'ecoulement uniforme}.
\newblock Technical report, EADS CCR, 2002.

\bibitem{delnevo2005aiaa}
A.~Delnevo, S.~Le~Saint, G.~Sylvand, and I.~Terrasse.
\newblock Numerical methods: Fast multipole method for shielding effects.
\newblock {\em 11th AIAA/CEAS Aeroacoustics Conference}, pages 2005--2971,
  2005.
\newblock Monterey.

\bibitem{DDMT}
F.~Dubois, E.~Duceau, F.~Mar\'echal, and I.~Terrasse.
\newblock Lorentz transform and staggered finite differences for advective
  acoustics.
\newblock Technical report, EADS and arXiv:1105.1458, 2002.

\bibitem{duprey}
S.~Duprey.
\newblock {\em Analyse Math\'ematique et Num\'erique du Rayonnement Acoustique
  des Turbor\'eacteurs}.
\newblock PhD thesis, EADS-CRC ; Institut Elie Cartan-Universit\'e Poincar\'e
  Nancy, 2005.

\bibitem{Ev73}
W.~Eversman.
\newblock Approximation for thin boundary layers in the sheared flow duct
  transmission problem.
\newblock {\em The Journal of the Acoustical Society of America},
  53(5):1346--1350, 1973.

\bibitem{2006:ParallelSolverLargeProblems}
L.~Giraud, J.~Langou, and G.~Sylvand.
\newblock On the parallel solution of large industrial wave propagation
  problems,.
\newblock {\em Journal of Computational Acoustics}, 14:83--111, 2006.

\bibitem{Glauert}
H.~Glauert.
\newblock The effect of compressibility on the lift of an aerofoil.
\newblock {\em Proceedings of the Royal Society of London. Series A, Containing
  Papers of a Mathematical and Physical Character}, 118(779):pp. 113--119,
  1928.

\bibitem{HWSD66}
F.~H. Harlow, J.~E. Welch, J.~P. Shannon, and B.~J. Daly.
\newblock {The MAC method}.
\newblock Technical report, Los Alamos Scientific Laboratory, march 1966.

\bibitem{JN80}
C.~Johnson and J.~C. N\'ed\'elec.
\newblock On the coupling of boundary integral and finite element methods.
\newblock {\em Mathematics of Computation}, 35(152):1063--1079, 1980.

\bibitem{PhDlegendre:2003}
G.~Legendre.
\newblock {\em {Rayonnement acoustique dans un fluide en {\'e}coulement :
  analyse math{\'e}matique et num{\'e}rique de l'{\'e}quation de Galbrun}}.
\newblock These, ENSTA ParisTech, September 2003.

\bibitem{Le91}
V.~Levillain.
\newblock {\em {Couplage \'el\'ements finis-\'equations int\'egrales pour la
  r\'esolution des \'equations de Maxwell en milieu h\'et\`erog\`ene}}.
\newblock PhD thesis, Th\`ese \'Ecole Polytechnique, 1991.

\bibitem{Li02}
S.~Lidoine.
\newblock {\em Approches th{\'e}oriques du probl{\`e}me du rayonnement
  acoustique par une entr{\'e}e d'air de turbor{\'e}acteur: Comparaisons entre
  diff{\'e}rentes m{\'e}thodes analytiques et num{\'e}riques}.
\newblock PhD thesis, Ecole centrale de Lyon, 2002.

\bibitem{mclean}
W.C.H. McLean.
\newblock {\em Strongly Elliptic Systems and Boundary Integral Equations}.
\newblock Cambridge University Press, 2000.

\bibitem{Morfey1971159}
C.L. Morfey.
\newblock Acoustic energy in non-uniform flows.
\newblock {\em Journal of Sound and Vibration}, 14(2):159 -- 170, 1971.

\bibitem{MI68}
P.~M. Morse and K.~Ingard.
\newblock {\em Theoretical acoustics}.
\newblock McGraw-Hill Book Co., 1968.

\bibitem{My80}
MK~Myers.
\newblock On the acoustic boundary condition in the presence of flow.
\newblock {\em Journal of Sound and Vibration}, 71(3):429--434, 1980.

\bibitem{Ne01}
J.C. N\'ed\'elec.
\newblock {\em {Acoustic and Electromagnetic Equations: Integral
  Representations for Harmonic Problems}}, volume 144 of {\em Applied Math.
  Sciences}.
\newblock Springer, 2001.

\bibitem{Pe75}
J.~P\'eriaux.
\newblock Three dimensional analysis of compressible potential flows with the
  finite element method.
\newblock {\em International Journal for Numerical Methods in Engineering},
  9(4):775--831, 1975.

\bibitem{Ra77}
J.~W.~S. Rayleigh.
\newblock {\em The theory of sound}.
\newblock Macmillan and co, London, 1877.

\bibitem{fmm}
V.~Rokhlin.
\newblock {Rapid solution of integral equations of classical potential theory}.
\newblock {\em Journal of Computational Physics}, 60:187--207, 1985.

\bibitem{introacous}
A.~Hirschberg S.~W.~Rienstra.
\newblock {\em An Introduction to Acoustics}.
\newblock Eindhoven University of Technology, 2004.

\bibitem{Sc68}
H.~A. Schenck.
\newblock {Improved Integral Formulation for Acoustic Radiation Problems}.
\newblock {\em The Journal of the Acoustical Society of America}, 44:41--58,
  1968.

\bibitem{So12}
A.~Sommerfeld.
\newblock {Die Greensche Funktionen der Schwingungsgleichung}.
\newblock {\em Jahresber. Deutsch. Math. Verein.}, 21:309--353, 1912.

\bibitem{sylvand2002phd}
G.~Sylvand.
\newblock {\em La m{\'e}thode multip{\^o}le rapide en
  {\'e}lectromagn{\'e}tisme. Performances, parall{\'e}lisation, applications}.
\newblock PhD thesis, Ecole des Ponts ParisTech, 2002.

\bibitem{Ta95}
A.~Taflove.
\newblock {\em {Computational Electrodynamics: The Finite-Difference
  Time-Domain Method.}}
\newblock Artech House, Norwood, MA, 1995.

\bibitem{Vi86}
J.~Virieux.
\newblock {PSV-wave propagation in heterogeneous media: velocity-stress finite
  difference method}.
\newblock {\em Geophysics}, 51:889--901, 1986.

\bibitem{wright}
M.C.M. Wright.
\newblock {\em Lecture Notes On The Mathematics Of Acoustics}.
\newblock Imperial College Press, 2005.

\bibitem{schur}
F.~Zhang.
\newblock {\em The Schur Complement and Its Applications}.
\newblock Numerical Methods and Algorithms. Springer, 2005.

\bibitem{Zi67}
O.~C. Zienkiewicz, D.~W. Kelly, and P.~Bettess.
\newblock The coupling of the finite element method and boundary solution
  procedures.
\newblock {\em International Journal for Numerical Methods in Engineering},
  11:355--375, 1977.

\end{thebibliography}

\end{document}